\documentclass[draftclsnofoot,12pt,onecolumn,journal]{IEEEtran}
\usepackage{cite}
\usepackage{amsmath}
\usepackage{amsfonts,amssymb}
\usepackage{mathrsfs}
\usepackage{graphicx,epsfig,epstopdf}
\usepackage[utf8]{inputenc}
\usepackage{CJK}
\usepackage{bm}
\usepackage{algorithm}
\usepackage{extarrows}
\usepackage{xcolor}
\usepackage{ragged2e}
\usepackage{array}
\newtheorem{remark}{\underline{Remark}}
\newtheorem{thr}{\underline{Theorem}}

\def\degree{${}^{\circ}$}
\newcommand{\tu}{\textup}

\newcommand{\diag}{\textup{diag}}
\def\st{\textup{s.t.}}

\begin{document}
\title{Beam Squint and Channel Estimation for \\
Wideband mmWave Massive \\
MIMO-OFDM Systems}

\author{Bolei Wang, Mengnan Jian, Feifei Gao,
Geoffrey Ye Li, \\ Shi Jin, and Hai Lin
\thanks{B. Wang, M. Jian, and F. Gao are with
Institute for Artificial Intelligence, Tsinghua University (THUAI),
State Key Lab of Intelligent Technologies and Systems, Tsinghua University,
Beijing National Research Center for Information Science and Technology (BNRist),
and Department of Automation, Tsinghua University,
Beijing, P.R.~China (e-mail:
\mbox{boleiwang@ieee.org};
jmn16@mails.tsinghua.edu.cn;
feifeigao@ieee.org).}
\thanks{G.~Y.~Li is with the School of Electrical and Computer Engineering, Georgia Institute of Technology, Atlanta, GA, USA (email: liye@ece.gatech.edu). }
\thanks{S. Jin is with the National Communications Research Laboratory, Southeast University, Nanjing 210096, P.R.~China (email: jinshi@seu.edu.cn).}
\thanks{H. Lin is with the Department of Electrical and Information Systems, Graduate School of Engineering, Osaka Prefecture University, Sakai, Osaka, Japan  (e-mail: hai.lin@ieee.org).}
\vspace{-15mm}
}
\maketitle
\begin{abstract}
With the increasing scale of antenna arrays in wideband millimeter-wave (mmWave) communications, the physical propagation delays of electromagnetic waves traveling across the whole array will become large and comparable to the time-domain sample period, which is known as the spatial-wideband effect.
In this case, different subcarriers in an orthogonal frequency division multiplexing (OFDM) system will ``see'' distinct angles of arrival (AoAs) for the same path.
This effect is known as \emph{beam squint}, resulting from the spatial-wideband effect, and makes the approaches based on the conventional multiple-input multiple-output (MIMO) model, such as channel estimation and precoding, inapplicable.
After discussing the relationship between beam squint and the spatial-wideband effect, we propose a channel estimation scheme for frequency-division duplex (FDD) mmWave massive MIMO-OFDM systems with hybrid analog/digital precoding, which takes the beam squint effect into consideration.
A super-resolution compressed sensing approach is developed to extract the frequency-insensitive parameters of each uplink channel path, i.e., the AoA and the time delay, and the frequency-sensitive parameter, i.e., the complex channel gain.
With the help of the reciprocity of these frequency-insensitive parameters in FDD systems, the downlink channel estimation can be greatly simplified, where only limited pilots are needed to obtain downlink complex gains and reconstruct downlink channels.
Furthermore, the uplink and downlink channel covariance matrices can be constructed from these frequency-insensitive channel parameters rather than through a long-term average,
which enables the minimum mean-squared error (MMSE) channel estimation
to further enhance performance.
Numerical results demonstrate the superiority of the proposed scheme over the conventional methods under general system configurations in mmWave communications.
\end{abstract}
\begin{IEEEkeywords}
Beam squint, wideband, millimeter wave, massive MIMO, mmWave, channel estimation, channel covariance reconstruction, angle reciprocity, delay reciprocity, hybrid precoding.
\end{IEEEkeywords}
\IEEEpeerreviewmaketitle
\section{Introduction}

Millimeter-wave (mmWave) communications have been widely recognized as a promising technology for future wireless networks
\cite{Heath-Mag-2014-5G,Heath-JSTSP-overview,Rappaport-proc-2014,SFW-BS-GC}.
With very wide frequency bands, mmWave communications can
offer unprecedented gigabits-per-second data rates and satisfy the rapidly growing demand of wireless traffic, such as in dynamic micro-cell or pico-cell (IEEE 802.11ad) systems \cite{CE-ref4-narrowband-atheta-Yuanjinghong-JSAC17}.
However, radio signals in mmWave bands suffer from the severe path loss  and are hard to bypass obstacles due to their weak diffractive ability \cite{severepathloss}.
To address this issue, massive multiple-input multiple-output (MIMO) technology is applied
to combat path loss, which meanwhile improves spectral and energy efficiencies \cite{GAO1}, as well as facilitates the exploitation of channel sparsity for mmWave communications.

In recent years, tremendous efforts have been devoted to applying massive MIMO in mmWave communications. With plenty of available spatial degrees of freedom in massive MIMO, only low computational linear precoding schemes, such as maximal ratio combining and zero-forcing,
are needed to mitigate inter-user interference and to achieve high data rates \cite{CE-ref4-narrowband-atheta-Yuanjinghong-JSAC17,
Marzetta-2010,Heath-multiuser-precoding-atheta}.
Since channel state information is crucial for massive MIMO systems, various channel estimation techniques \cite{CE1-narrowband,CE2-narrowband-atheta,CE3-narrowband-atheta,
CE4-narrowband-atheta-Fang-tensor,CE5-wideband-atheta-Fang-tensor-simu,
CE-ref1-narrowband-atheta-Fang-TWC18,
CE-ref2-wideband-atheta-Hanzhu-JSAC17,
CE-ref3-narrowband-atheta-Heath-TWC17,
CE-ref4-narrowband-atheta-Yuanjinghong-JSAC17,
DFT-Xie}
have been developed for mmWave communications to exploit channel sparsity in angle domain and delay domain.
A minimum mean-squared error (MMSE) estimator has been proposed in \cite{CE1-narrowband} with the help of channel covariance.
In \cite{CE3-narrowband-atheta,CE2-narrowband-atheta}, channel estimation is transformed to a sparse signal recovery problem by exploiting sparse scattering property of mmWave channels. Super-resolution algorithms have also been developed in
\cite{CE4-narrowband-atheta-Fang-tensor,
CE5-wideband-atheta-Fang-tensor-simu,
CE-ref1-narrowband-atheta-Fang-TWC18,
DFT-Xie,SFW-BS-GC}
to eliminate the grid mismatch in dictionary-based on-grid approaches.

However, in a system with large-scale antenna arrays, different antennas may receive different time-domain symbols from the same physical path at the same sampling time due to the large propagation delay of electromagnetic waves travelling across the whole antenna array, which is known as the \emph{spatial-wideband effect} \cite{SFW-APCC,SFW-TSP,SFW-CM}.
In this case, the massive MIMO channel model,
which only considers phase difference and ignores delay difference among the received signals at different antennas,
are not applicable any more. The algorithms based on such models, such as for channel estimation and precoding, need to be revised.

The spatial-wideband effect causes \emph{beam squint} in the frequency domain.
As will be shown in Section II, the level of the beam squint effect is proportional to that of the spatial-wideband effect. Hence, it only becomes remarkable and non-negligible in large-scale antenna arrays and/or broadband transmission.
The beam squint effect has been initially investigated in radar systems and array signal processing \cite{squint1,squint2,squint3}
because radar systems have employed large-scale antenna arrays at the earliest,
dating back to the late 1950s \cite{radar1950}.
In massive MIMO communications
\cite{SFW-BS-GC,SFW-TSP,SFW-CM,Cai-GC2016,Geoffrey-Subarray-JSAC,Geoffrey-Subarray-CM},
beam squint renders the observed angles of arrival (AoAs) and the anticipant angles of departure (AoDs) frequency-dependent.
Specifically, in an OFDM system,
beam squint makes different subcarriers to observe distinct AoAs for the same physical path.
Conversely, if one ignores beam squint and deploys the identical beam-steering vector at different subcarriers, then signals at different subcarriers will point towards different physical directions.
Therefore, the beam squint effect should be carefully considered in channel estimation, especially for the physical angles-based approaches.

Another challenge caused by beam squint is in downlink channel estimation and precoding for FDD systems with hybrid transceivers. Since the analog precoder, i.e., the phase setup in phase shifters, is fixed during one OFDM block, it cannot generate the frequency-dependent steering vectors, which, however, is obligatory when considering the beam squint effect. To address this issue, we propose to utilize several RF chains to cooperatively generate frequency-dependent steering vectors for each path via tuning the digital precoder for each subcarrier, which, to the best of our knowledge, is the first work to address this issue and will be discussed in Section~\ref{sec:dl}.

As mmWave communications highly rely on the precise alignment of beams between the transmitter and the receiver, beam squint will result in severe performance degradation if not carefully treated.
A combining pattern for high-dimensional receivers with beam squint has been investigated in \cite{Sayeed-ICC15}
for mmWave MIMO channels in the line-of-sight (LoS) scenario.
The algorithms in \cite{Cai-GC2016,Cai-ISIT2017} partly compensate for beam squint in analog phased arrays by increasing the codebook size in beamforming.
Based on the massive MIMO-OFDM channel model considering the spatial-wideband effect, we have developed
the fast Fourier transform-based channel estimation approach \cite{SFW-TSP}, which can address the beam squint issue in wideband communications but encounters error floors at high SNR regions.
For the single-user mmWave systems with hybrid transceivers, a compressive sensing based channel estimation technique has been proposed in \cite{Nuria_GC18} to estimate uplink channels with beam squint effect.

In this paper, we investigate both uplink and downlink channel estimations for multi-user mmWave massive MIMO-OFDM systems with hybrid precoding, which takes the beam squint effect into consideration.
The channel is depicted as a function of physical parameters, including the frequency-insensitive ones, i.e., the angle of arrival/departure (AoA/AoD) and the time delay of each path, as well as the frequency-sensitive one, i.e., the complex gain.
A super-resolution compressed sensing algorithm is proposed with the adaptive-updating dictionary to extract the physical parameters from the uplink channel estimation.
With the help of the reciprocity of the frequency-insensitive parameters,
downlink channel estimation for frequency-division duplex (FDD) systems can be greatly simplified and only a small amount of training and user feedback are needed.
Moreover, the channel covariance matrices for both uplink and downlink channels can be reconstructed from these parameters rather than through the long-term average,
which facilitates MMSE channel estimation to further enhance the performance.
Our numerical results corroborate our theoretical analysis and demonstrate the superiority of the proposed scheme over the existing algorithms under general mmWave system configurations.

The rest of this paper is organized as follows.
Section~II proposes the wideband channel model for mmWave massive MIMO-OFDM systems and its characteristics.
Section~III introduces the system model with hybrid precoding scheme.
Section~IV derives a super-resolution compressive sensing based approach to extract initial parameters of uplink channels.
Sections~V and VI develop an efficient multi-user channel estimation strategy with limited feedback for FDD systems.
Simulation results are provided in Section VII and
Section~VIII concludes this paper.

\textbf{Notations: }
Uppercase and lowercase boldface denote matrices and vectors, respectively.
Superscripts $(\cdot)^T,(\cdot)^H,(\cdot)^*,(\cdot)^\dag$
stand for the transpose, the conjugate-transpose, the conjugate, and the pseudo-inversion of a matrix or a vector, respectively.
Symbols $\mathbf I$, $\mathbf 1$, and $\mathbf 0$ represent
the identity matrix, the all-ones matrix, and the all-zeros matrix
while their subscripts, if needed, indicate the dimensionality.
Symbols $\odot$ and $\otimes$ denote the Hadamard product and Kronecker product of two matrices, respectively.
$\mathbb E\{\cdot\}$ denotes the expectation and $tr(\cdot)$ represents the matrix trace operation.
$\|\mathbf a\|_2$ and $\|\mathbf A\|_F$ denote the Euclidean norm of vector $\mathbf a$ and the Frobenius norm of matrix $\mathbf A$, respectively.
We use $[\mathbf A]_{m,n}$, $[\mathbf A]_{:,n}$, and $[\mathbf A]_{m,:}$ to denote the $(m,n)$th element, the $n$th column, and the $m$th row of matrix $\mathbf A$, respectively.
$|\mathcal A|$ is the cardinality of set $\mathcal A$.
$\textup{diag}\{\mathbf a\}$ denotes the diagonal matrix comprising vector $\mathbf a$'s elements
and $\textup{diag}\{\mathbf A\}$ represents the column vector extracted from the diagonal entries of matrix $\mathbf A$.

\section{Beam Squint in Wideband Massive MIMO Systems}
\label{sec:channel}

Consider a mmWave massive MIMO-OFDM system with a base station (BS) and $K$ single-antenna users randomly distributed throughout the cell. The BS is equipped with an $M$-antenna uniform linear array (ULA) and the antenna spacing for the BS is $d$.
Orthogonal frequency-division multiplexing (OFDM) with $N_c$ subcarriers  is adopted for combating the multipath delay spread.
If the overall transmission bandwidth is $W$, then the subcarrier spacing will be $\eta = W/N_c$.

\subsection{Wideband mmWave Massive MIMO-OFDM Channel Model}

Suppose there are $L_k$ incident paths from the $k$th user to the BS.
Denote $\tau _{k,l,m}$ as the time delay of the $l$th path from the $k$th user to the $m$th antenna of the BS
and denote $\tau_{k,l} \triangleq \tau_{k,l,1}$ for notational simplicity,
where $k\in\{ 1,\dots,K\}$,
$l\in\{ 1,\dots,L_k\}$,
and $m\in\{ 1,\dots,M\}$.
Denote $\vartheta_{k,l}$ as the AoA of the $l$th path from the $k$th user and define
$\psi _{k,l} \triangleq \frac{d \sin \vartheta_{k,l} }{\lambda _c} $ as the normalized AoA,
where $\lambda _c$ is the carrier wavelength.
Then, based on the far-field assumption \cite{Tse-book04} that the antenna array sizes are much smaller than the distance between the transmitter and the receiver,
\begin{align}
\tau_{k,l,m} & = \tau_{k,l} + (m-1) \frac{d \sin \vartheta_{k,l} }{c}
= \tau_{k,l} + (m-1) \frac{ \psi_{k,l} }{f_c},
\label{tau-plmi}
\end{align}
where $c$ is the speed of light
and $f_c = c / \lambda_c$ is the carrier frequency.

Denote the complex channel gain of the $l$th path from the $k$th user as $\bar{\alpha} _{k,l}$. Then, the impulse response of the uplink channel between the $m$th antenna at the BS and the $k$th user can be expressed as
\begin{align}
h_{k,m}^T (t) &= \sum_{l=1}^{L_k} \bar{\alpha} _{k,l}
  e^{-j2\pi f_c \tau_{k,l,m} }
  \delta ( t - \tau_{k,l,m} )
= \sum_{l=1}^{L_k}
  \alpha _{k,l} e^{-j2\pi (m-1) \psi _{k,l} }
  \delta ( t - \tau_{k,l,m} )    ,
\label{ht}
\end{align}
where
$\alpha _{k,l} \triangleq \bar{\alpha} _{k,l} e^{-j2\pi f_c \tau_{k,l} }$ is the equivalent complex gain.

By taking the Fourier transform of \eqref{ht}, the \emph{frequency response} between the $m$th antenna at the BS and the $k$th user can be obtain as
\begin{align}
h_{k,m}^F (f) &
= \sum_{l=1}^{L_k} \alpha _{k,l} e^{-j2\pi (m-1) \psi _{k,l} }
  e^{-j2\pi f \tau_{k,l,m}}
= \sum_{l=1}^{L_k} \alpha _{k,l}
  e^{-j2\pi (m-1) \psi _{k,l} (1+\frac{f}{f_c})}
  e^{-j2\pi f \tau_{k,l}}  ,
\label{hf}
\end{align}
where the second equality utilizes the result in \eqref{tau-plmi}.

Denote
\begin{align}
\Xi _{k,l} (f)  \triangleq  \left(1+\frac{f}{f_c}\right) \psi _{k,l}
.
\end{align}
Stacking all $h_{k,m}^F (f)$'s from different antennas into a vector yields
\begin{align}
\mathbf h_k^F (f) & \triangleq \sum_{l=1}^{L_k} \alpha _{k,l}
  \mathbf a (\Xi _{k,l} (f))
  e^{-j2\pi f \tau_{k,l}}   ,
\label{Hf}
\end{align}
where
\begin{align}
  & \mathbf a (\Xi _{k,l} (f))  \triangleq  [1,e^{-j2\pi \Xi _{k,l} (f)},\dots,e^{-j2\pi (M-1)\Xi _{k,l} (f)} ]^T
    \in \mathbb C^{M\times 1}
\label{axif}
\end{align}
is the \emph{spatial-domain steering vector}.

The proposed model in \eqref{Hf} accurately depicts the wideband massive MIMO-OFDM channel.
Different from the widely-used mmWave models,
the \emph{steering vectors} in \eqref{axif} is \emph{frequency-dependent},
which is referred to as the \emph{beam squint} effect.

\begin{remark}
   For the widely-used MIMO channel model, the spatial-domain steering vector is independent of frequency,
   which is true when the difference of time delays between different antennas are negligible, i.e., $
       \tau_{k,l,m} \approx \tau_{k,l}, \
       \forall m\in\{ 1,\dots,M\}
       $.
   However, when both the number of antennas in one dimension and the transmission bandwidth become large,
   $\max (\tau_{k,l,m} - \tau_{k,l}) \ll T_s $ or
   $\max (\tau_{k,l,m} - \tau_{k,l}) \ll 1/W$ does not hold any more
   and $(m-1) \frac{d \sin \vartheta_{k,l} }{c}$ in \eqref{tau-plmi}
   cannot be ignored.
   For example, a 128-antenna ULA with half wavelength antenna spacing can induce a maximum delay of $1.36 T_s$ and result in frequency-dependent steering vectors when operating at 28~GHz with 600~MHz bandwidth.
\end{remark}

\begin{remark}
The derived channel model in \eqref{Hf} with beam squint can be straightforwardly extended to multi-dimensional setups.
For example, consider that users are also equipped with large $M_U$-antenna ULAs with antenna spacing $d_U$.
Denote $\vartheta _{k,l}$, $\vartheta _{U,k,l}$ as the AoA and the AoD corresponding to the $l$th path from the $k$th user and
similarly, denote
$\psi _{k,l} \triangleq \frac{d \sin \vartheta_{k,l} }{\lambda _c} $ and $\psi_{U,k,l} \triangleq  \frac{d_U \sin \vartheta_{U,k,l}  }{\lambda _c} $ as the normalized AoA and AoD, respectively.
By the similar mathematical manipulation,
the channel of the $k$th user can be arranged by a matrix as
\begin{align}
\mathbf H_k^F (f) & = \sum_{l=1}^{L_k} \alpha _{k,l}
  \mathbf a (\Xi _{k,l} (f)) \mathbf a_U^H (\Xi _{U,k,l} (f))
  e^{-j2\pi f \tau_{k,l}}  ,
\label{Hf2}
\end{align}
where
$
  \mathbf a (x) \triangleq  [1,e^{-j2\pi x},\dots,e^{-j2\pi (M-1)x} ]^T
    \in \mathbb C^{M\times 1}
$
and
$
  \mathbf a_U (x) \triangleq  [1,e^{-j2\pi x},\dots,e^{-j2\pi (M_U-1)x} ]^T
    \in \mathbb C^{M_U\times 1}
$
are the \emph{spatial-domain steering vectors} of the BS and the user, respectively, with
$  \Xi _{k,l} (f)  \triangleq  (1+\frac{f}{f_c}) \psi _{k,l} $
and
$  \Xi _{U,k,l} (f)  \triangleq  (1+\frac{f}{f_c}) \psi _{U,k,l} $.
The $(m,i)$th element of $\mathbf H_k^F (f)$ in \eqref{Hf2} denotes the frequency response between the $m$th antenna of the BS and the $i$th antenna of the $k$th user.
\end{remark}

\subsection{Beam Squint over OFDM Subcarriers}

In this subsection, we discuss how the frequency-dependent steering vectors interacts in the OFDM modulation.
From \eqref{Hf}, the channel between the BS and the $k$th user at the $q$th subcarrier can be expressed by
\begin{align}
\mathbf h_{k,q} & = \sum_{l=1}^{L_k}
  \alpha _{k,l} \mathbf a (\Xi _{k,l} ((q-1)\eta))
  e^{-j2\pi (q-1) \eta \tau_{k,l}} ,~q \in \{ 1,\dots,N_{c}\} .
\label{hkq}
\end{align}

After transforming \eqref{hkq} to virtual angle domain
\cite{DFT-Xie,SFW-TSP}
by discrete Fourier transform (DFT),
we can obtain the following theorem, which demonstrates the beam squint effect in wideband mmWave massive MIMO-OFDM systems.

\begin{thr}
  The spatial-wideband effect induces each path in angle domain to squint along with subcarrier indices.
  The maximum squint along the angular indices is approximately the propagation delay across the antenna array in sample periods.
\end{thr}

\begin{IEEEproof}
The channel in \eqref{hkq} in virtual angle domain can be computed as
\begin{align}
[\mathbf F_M^H \mathbf h_{k,q}]_{v} & =
\frac{1}{\sqrt M}
\sum_{l=1}^{L_k} \alpha_{k,l} e^{-j2\pi (q-1) \eta \tau_{k,l}}
\times \notag\\
& \phantom {mmmmmm}\frac{\sin(\pi M  [\psi_{k,l} (1+(q-1)\eta/f_c ) - \frac{v}{M}])}
{\sin (\pi  [\psi_{k,l} (1+(q-1)\eta/f_c ) - \frac{v}{M}])}
e^{-j\pi (M-1) [\psi_{k,l} (1+(q-1)\eta/f_c ) - \frac{v}{M}]}
,
\label{FMh}
\end{align}
where $\mathbf F_M$ is the $M$-dimensional normalized DFT matrix.
As $M$ is large, the values of the function
$\left|\frac{\sin(\pi M x)}{\sin (\pi x)} \right|$
are significant only when
$x \simeq 0$,
which indicates that the power of the $l$th path concentrates on angular index
$v_{l,q} \simeq M \psi_{k,l} ( 1+(q-1)\eta/f_c )$ and squints along with subcarrier index $q$.
For the $l$th path, the squint over all subcarriers can be expressed as
\begin{align}
|v_{l,N_c} - v_{l,1}| & =
M \psi_{k,l} \frac{(N_c-1) \eta}{f_c}
= \left(M \frac{d\sin \vartheta_{k,l}}{\lambda_c f_c} \right) W
\simeq \tau_{k,l}^{prop}W = \frac{\tau_{k,l}^{prop} }{T_s}
,
\label{beam_squint_index}
\end{align}
where $\tau_{k,l}^{prop} \triangleq (M-1) \frac{d\sin \vartheta_{k,l}}{c}$ is the physical propagation delay of the $l$th path across the whole antenna array.
It can be further verified that the three ``approximate equals'' above will turn into the ``strict equals'' as $M \to \infty$.
\end{IEEEproof}

\begin{figure}
  \centering
  \includegraphics[width=150mm]{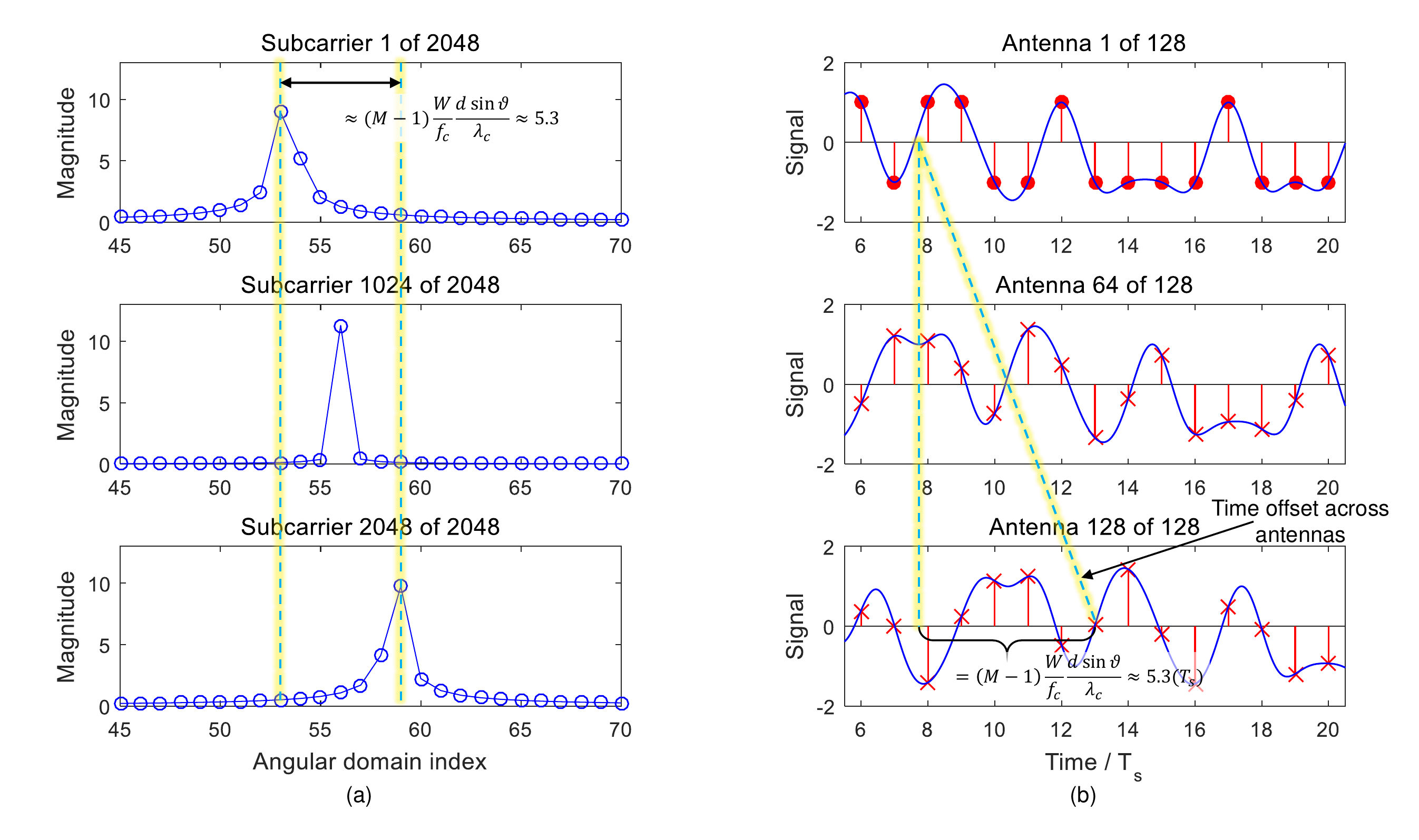}\\
  \caption{A one-path channel in virtual angle domain at different subcarriers, with AoA $\vartheta = 56.5^\circ$, $M=128$, $N_c = 2048$, $d / \lambda_c = 0.5$, $W/f_c = 0.1$.}
  \label{fig_squint1}
\end{figure}

Theorem 1 clarifies the relationship between the beam squint effect and the spatial-wideband effect.
Fig.~\ref{fig_squint1} illustrates a one-path channel in virtual angle domains at different subcarriers, where the BS ``sees'' different angles of a certain path at different subcarriers. For multi-dimensional setups as indicated in remark 2, theorem 1 can be applied to each dimension to observe the beam squint level of each dimension.
Nevertheless, the beam squint effect is hard to be observed and is rarely discussed in the conventional small MIMO communication systems, where the propagation delay across antennas is small and thus the spatial-wideband effect can be neglected.

\section{System Model with Spatial-Wideband Effect}
\label{sec:system}

In section~\ref{sec:channel}, we derive the wideband mmWave massive MIMO channel model with consideration of the spatial-wideband effect, of which the manifestation in frequency domain is the beam squint effect. It should be noted that this model is related to the array manifold and irrelevant to the architecture behind the array. Therefore, the full-digital and the hybrid analog/digital precoding systems share the same channel model in \eqref{hkq}. In the sequel, we consider the mmWave systems under the phase shifter-based hybrid architecture, as it is much more practical in mmWave communications.

\begin{figure}
  \centering
  \includegraphics[width=90mm]{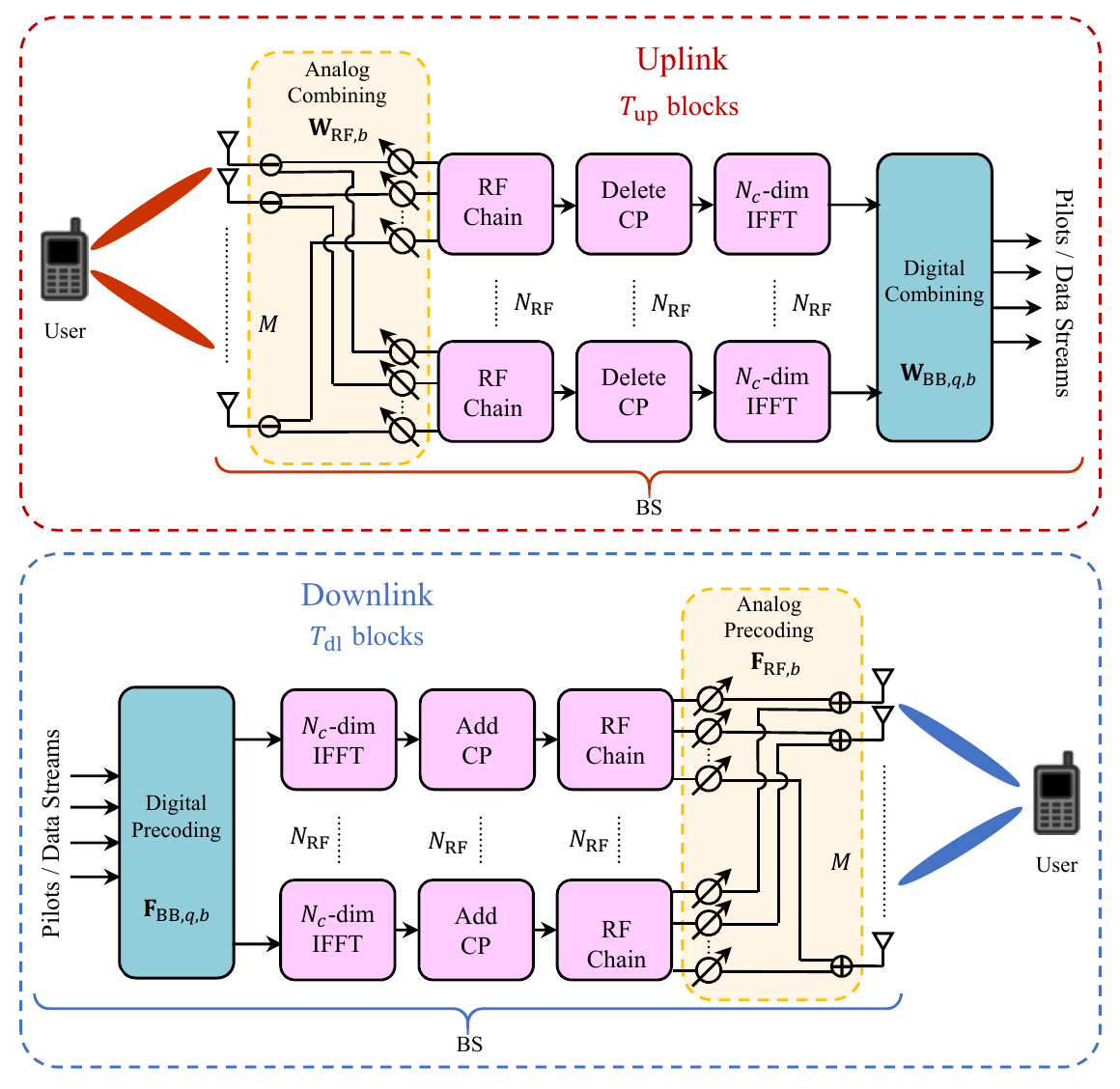}\\
  \caption{The block diagram of a user and the BS with hybrid precoding.}
  \label{figsystem}
\end{figure}
Following the system setup in Section~\ref{sec:channel}, assume the BS to have $N_{\textup{RF}}$ radio frequency (RF) chains. We employ $T_{\textup{up}}$ successive OFDM blocks for uplink channel estimation.
The hybrid precoder/combiner at the BS for the $q$th subcarrier and the $b$th block can then be denoted as
$\mathbf W_{q,b} = \mathbf W_{\textup{RF},b} \mathbf W_{\textup{BB},q,b} \in \mathbb C^{M \times N_{\textup{RF}} }$,
where $\mathbf W_{\textup{RF},b} \in \mathbb C^{M \times N_{\textup{RF}} } $ is the analog combiner implemented by phase shifters at the $b$th block and $\mathbf W_{\textup{BB},q,b} $ is the digital baseband combiner at the $q$th subcarrier and the $b$th block.
The system architecture is illustrated in Fig.~\ref{figsystem},
where the downlink channel model will be discussed in Section~\ref{sec:dl}.

Assume that $P$ of $N_c$ subcarriers are exclusively assigned to each user as pilots and the set of pilot subcarrier indices for the $k$th user is denoted by
$\mathcal P_k = \{p_{k,1},\dots,p_{k,P}\}$.
For the $k$th user, the received signal vector at the BS in the $q$th subcarrier at $M$ antennas in the $b$th block can be expressed as
\begin{align}
\mathbf y_{k,q,b} = \mathbf W_{q,b}^H \mathbf h_{k,q} x_{k,q,b} + \mathbf W_{q,b}^H \mathbf n_{k,q,b}
, \quad
q \in \mathcal P_k
,
\end{align}
where $x_{k,q,b}$ is the pilot symbol from the $k$th user at the $q$th subcarrier in the $b$th block and $\mathbf n_{k,q,b} \in \mathbb C^{M\times 1}$ is the corresponding additive Gaussian noise with each element independently distributed as
$\mathcal {CN} (0, \sigma _n^2)$.

Note that the pilot symbols are known at both the BS and users. Stacking the received pilots of $T_{\textup{up}}$ blocks into a vector, we have
\begin{align}
\mathbf y_{k,q} \triangleq  \bigg[
\frac{1}{x_{k,q,1}} \mathbf y_{k,q,1}^T ,\dots, \frac{1}{x_{k,q,b}} \mathbf y_{k,q,T_{\textup{up}}}^T
\bigg]^T =
\mathbf W_q^H \mathbf h_{k,q} + \tilde{\mathbf W}_q^H \tilde{\mathbf n}_{k,q}
,
\end{align}
where $\mathbf W_q \triangleq \big[
\mathbf W_{q,1} ,\dots, \mathbf W_{q,T_{\textup{up}}}
\big] \in \mathbb C^{M \times N_{\textup{RF}} T_{\textup{up}}}$,
\begin{align}
\tilde{\mathbf W}_q & \triangleq
\left[
\begin{array}{cccc}
\mathbf W_{q,1} & \mathbf 0 & \cdots & \mathbf 0 \\
\mathbf 0 & \mathbf W_{q,2}  & \cdots & \mathbf 0 \\
\vdots & \vdots & \ddots & \vdots \\
\mathbf 0 & \mathbf 0 & \cdots & \mathbf W_{q,T_{\textup{up}}}
\end{array}
\right]
\in \mathbb C^{MT_{\textup{up}} \times N_{\textup{RF}} T_{\textup{up}}}
,
\label{Wtildeq}
\end{align}
and
$\tilde{\mathbf n}_{k,q} \triangleq \Big[
\frac{1}{x_{k,q,1}} \mathbf n_{k,q,1}^T ,\dots, \frac{1}{x_{k,q,b}} \mathbf n_{k,q,T_{\textup{up}}}^T
\Big]^T \in \mathbb C^{MT_{\textup{up}} \times 1 }$.
Denote $\mathbf h_k \triangleq \big[
\mathbf h_{k, p_{k,1}}^T ,\dots, \mathbf h_{k, p_{k,P}}^T
\big]^T \in \mathbb C^{MP \times 1}$.
Collecting $\mathbf y_{k,q} $ at different subcarriers, we have
\begin{align}
\mathbf y_k \triangleq \big[
\mathbf y_{k, p_{k,1}}^T ,\dots, \mathbf y_{k, p_{k,P}}^T
\big]^T =
\mathbf W^H \mathbf h_k + \mathbf n_k
,
\label{yk}
\end{align}
where
\begin{align}
{\mathbf W} & \triangleq
\left[
\begin{array}{cccc}
\mathbf W_{p_{k,1}} & \mathbf 0 & \cdots & \mathbf 0 \\
\mathbf 0 & \mathbf W_{p_{k,2}}  & \cdots & \mathbf 0 \\
\vdots & \vdots & \ddots & \vdots \\
\mathbf 0 & \mathbf 0 & \cdots & \mathbf W_{p_{k,P}}
\end{array}
\right]
\in \mathbb C^{MP \times N_{\textup{RF}} P T_{\textup{up}}}
\notag
\end{align}
and $
{\mathbf n_k} \triangleq
\Big[
\big( \tilde{\mathbf W}_{p_{k,1}}^H \tilde{\mathbf n}_{p_{k,1}} \big)^T ,\dots,
\big( \tilde{\mathbf W}_{p_{k,P}}^H \tilde{\mathbf n}_{p_{k,P}} \big)^T
\Big]^T
\in \mathbb C^{ N_{\textup{RF}} P T_{\textup{up}} \times 1}
$.

From \eqref{hkq}, $\mathbf h_k$ can be written as
\begin{align}
{\mathbf h}_k & =
\sum_{l=1}^{L_k} \alpha_{k,l} {\mathbf p_k}(\psi_{k,l}, \tau_{k,l})
,
\label{hktilde}
\end{align}
where
\begin{align}
&{\mathbf p_k}(\psi_{k,l}, \tau_{k,l}) \triangleq \Big[
\mathbf a^T (\Xi _{k,l} ((p_{k,1}-1)\eta)) e^{-j2\pi (p_{k,1}-1) \eta \tau_{k,l}},
\notag \\ &\kern 100pt \dots,
\mathbf a^T (\Xi _{k,l} ((p_{k,P}-1)\eta)) e^{-j2\pi (p_{k,P}-1) \eta \tau_{k,l}}
\Big]^T
\in \mathbb C^{MP \times 1}
\label{p_t}
\end{align}
can be regarded as the channel basis for user $k$ with respect to AoA $\psi _{k,l}$ and path delay $\tau _{k,l}$.
Denote
$\bm \alpha_k \triangleq \left[
\alpha_{k,1}  ,\dots, \alpha_{k,L_k}
\right]^T \in \mathbb C^{L_k \times 1}$,
$\bm \psi_k \triangleq \left[
\psi_{k,1}  ,\dots, \psi_{k,L_k}
\right]^T \in \mathbb C^{L_k \times 1}$ ,
and $\bm \tau_k \triangleq \left[
\tau_{k,1}  ,\dots, \tau_{k,L_k}
\right]^T \in \mathbb C^{L_k \times 1}$.
Equation \eqref{hktilde} can be expressed in the vector/matrix form as
\begin{align}
{\mathbf h}_k & =
{\mathbf P}_k(\bm \psi_k, \bm \tau_k) \bm \alpha_k,
\label{hk}
\end{align}
where
\begin{align}
{\mathbf P}_k (\bm \psi_k, \bm \tau_k) \triangleq \left[
{\mathbf p}_k (\psi _{k,1} , \tau_{k,1}),\dots,
{\mathbf p}_k (\psi _{k,L_k} , \tau_{k,L_k})
\right]
\in \mathbb C^{MP \times L_k}
.
\label{P_t}
\end{align}
Equation \eqref{hk} provides a sparse representation of the wideband channel with the basis in \eqref{p_t} that considers the beam squint effect.

Similar to \cite{SFW-TSP}, it can be readily verified that
$\lim_{M,P \to \infty} \frac{1}{MP}
{\mathbf p}_k^H (\psi_1, \tau_1) {\mathbf p_k} (\psi_2, \tau_2) = \delta(\psi_1 - \psi_2) \delta(\tau_1 - \tau_2) $,
which we call the \emph{asymptotical angle-delay orthogonality}.
Assume that channel gains of different multipath components are with zero mean and independent with each other,
i.e.,
\begin{align}
\mathbb E\{ \bm \alpha _k \bm\alpha _k^H \} =
\diag \big\{ \mathbb E\{ | \alpha _{k,1} |^2 \} ,\dots, \mathbb E\{ | \alpha _{k,L_k} |^2 \}
\big\} \triangleq \bm \Lambda_k,
\label{Lambdak}
\end{align}
where $\mathbb E\{ | \alpha _{k,l} |^2 \}$ is the average power of the corresponding multipath component.
Based on the above discussion, the covariance matrix of the uplink channel for user $k$ can be expressed as
\begin{align}
{\mathbf R}_k^U \triangleq \mathbb E \Big\{ {\mathbf h}_k {\mathbf h}_k^H \Big\} =
{\mathbf P_k} (\bm \psi_k, \bm \tau_k)
\bm \Lambda_k
{\mathbf P}_k^H (\bm \psi_k, \bm \tau_k)
\in \mathbb C^{MP \times MP}
.
\label{Rku}
\end{align}
As $\tu{rank}({\mathbf R}_k^U) \leq \tu{rank}(\bm \Lambda_k) = L_k \ll MP$,
${\mathbf R}_k^U $ is a pretty low-rank matrix.
Since
$\frac{1}{\sqrt{MP}}{\mathbf P_k} (\bm \psi_k, \bm \tau_k)$ is a tall matrix with asymptotically mutually-orthogonal columns of unit length,
\begin{align}
{\mathbf R}_k^U =
\bigg( \frac{1}{\sqrt{MP}}
{\mathbf P_k} (\bm \psi_k, \bm \tau_k) \bigg)
\bigg( MP \bm \Lambda_k \bigg)
\bigg( \frac{1}{\sqrt{MP}}
{\mathbf P}_k^H (\bm \psi_k, \bm \tau_k) \bigg)
\label{Rkue}
\end{align}
provides a good approximation of eigenvalue decomposition.

With the aid of the \emph{AoA-delay reciprocity}
\cite{SFW-TSP,SFW-CM}, downlink channel covariance matrices can also be reconstructed via the physical parameters of uplink channels, which will be discussed in Section~\ref{sec:dl}.

\section{Initial Uplink Channel Parameter Extraction}
\label{sec:initial}

At the very beginning, all users stay in the dark as far as the BS is concerned and thus
orthogonal trainings have to be applied to avoid the inter-user interference and pilot contamination at the BS. In this case, we operate with the frequency orthogonality among difference users
\cite{OFDMA-TWC,SFW-TSP}
by setting
$\mathcal P_k \cap \mathcal P_r = \varnothing, $
$\forall k\neq r$,
at this very beginning phase.
During this phase, the initial AoA, time delay, and complex gain of each path are estimated for all users and we call this phase \emph{initial parameter extraction}.

In this section, we introduce a parameter extraction algorithm for this stage, which suffices for the subsequent multi-user uplink and downlink channel estimations, as will be shown in Sections~V and VI.

\subsection{Problem Formulation}

For the $k$th user, the uplink pilots transmission process is given in \eqref{yk}.
Substituting \eqref{hk} into \eqref{yk}, we have
\begin{align}
\mathbf y_k =
\mathbf W^H {\mathbf P}_k(\bm \psi_k, \bm \tau_k) \bm \alpha_k + \mathbf n_k
.
\label{yk2}
\end{align}
Our goal is to extract
the physical parameters, $\{ \bm \psi_k, \bm \tau_k, \bm \alpha _k \}$, from $\mathbf y_k$.
As the number of the parameters is generally far fewer than the dimensionality of $\mathbf y_k$, i.e., $3L_k \ll N_{\textup{RF}} P T_{\textup{up}}$, the compressive sensing could be a powerful tool to solve this parameter extraction problem.

\subsection{Algorithm Derivation}
\label{sec:alg1}

Compressive sensing \cite{Geoffrey-CS} can be used in mmWave massive MIMO channel estimation.
Many existing techniques
discretize
$\bm \psi_k$ and $\bm \tau_k$  into  finite sets of grid points
and then transform the parameters extraction into a sparse recovery problem
$\mathbf y_k = \mathbf A_k^{grid} \bm \alpha _k^{grid}$, where
$\mathbf A_k^{grid} \in \mathbb C^{N_{\textup{RF}} P T_{\textup{up}} \times N_{grid}} $ acts as the overcomplete dictionary constructed by a series of $N_{grid}$ possible discretized values of $\{ \bm \psi_k, \bm \tau_k\}$.
Since $\{ \bm \psi_k, \bm \tau_k\}$ corresponds to continuous-valued AoAs and path delays in the real world, these on-grid approaches will
suffer from performance degradation when a $\{ \bm \psi_k, \bm \tau_k\}$ does not fall into the pre-defined grid points, called the \emph{grid mismatch} \cite{offgrid}.
Grid mismatch can be alleviated by increasing the size of grid dictionary, which however will increase computational complexity.

To obtain the physical parameters, we propose a compressive sensing-based \emph{gridless (off-grid)} approach instead, where
the dictionary is not pre-defined and remains unknown during the iterative parameter extraction.
The number of channel paths is initialized to a relatively large value, denoted as $L_M ~ (\geq L_k)$.
The problem can be formulated as
\begin{align}
\min_{\bm \psi, \bm \tau, \bm \beta }  & \quad
\| \bm \beta  \| _0 \notag\\
& \st \quad \|
\mathbf y_k - \mathbf W^H \mathbf P_k (\bm \psi, \bm \tau)  \bm \beta
\| _2 \leq \xi
,
\label{op0}
\end{align}
where $\| \bm \beta \| _0 $ stands for the number of nonzero entries of vector $\bm \beta $ and the small positive number, $\xi$, controls the error tolerance related to the noise statistics.

For the optimal $\bm\beta$ in \eqref{op0},
its dimension is expected to reduce to the number of real paths, $L_k$, and it accordingly converges to the real channel gain vector, $\bm \alpha_k$.
Different from the existing algorithms,
the uplink channel is expanded with the basis in \eqref{p_t}, which considers the beam squint effect, consists of frequency-dependent steering vectors, and ensures the channel parameters to be accurately extracted.

Similar to \cite{log-sum-1},
to address the NP-hard optimization problem in \eqref{op0},
the log-sum sparsity-encouraging function
\begin{align}
J_0 (\bm \beta) \triangleq  \sum_{l=1}^{L_k}
\log \big(\big| [\bm \beta]_l \big| ^2 + \epsilon \big)
\label{J0}
\end{align}
can be applied,
where
the concrete value of $\epsilon$ will be discussed in the next subsection.
To address the constraint in \eqref{op0}, the data fitting term
$\lambda \|
\mathbf y_k - \mathbf W^H \mathbf P_k (\bm \psi, \bm \tau)  \bm \beta
\|_2^2 $
is included in the cost function.
Then, the optimization problem in \eqref{op0} is transformed into
\begin{align}
\min_{\bm \psi, \bm \tau, \bm \beta }
 \quad
\underbrace{
\sum_{l=1}^{L_k}
\log \big(\big| [\bm \beta]_l \big| ^2 + \epsilon \big)
+
\lambda \|
\mathbf y_k - \mathbf W^H \mathbf P_k (\bm \psi, \bm \tau)  \bm \beta
\|_2^2
}_{J_{\lambda} (\bm \psi, \bm \tau, \bm \beta)}
,
\label{opJ}
\end{align}
where  $\lambda > 0$ is the regularization parameter
and $J_{\lambda} (\bm \psi, \bm \tau, \bm \beta)$ is defined as the corresponding optimization objective.

Denote the estimates of the complex gain, the AoA, and the path delay at the $n$th iteration as $\bm \beta^{(n)} $, $\bm \psi^{(n)} $, and $\bm \tau^{(n)} $, respectively.
Utilizing the majorization-minorization (MM) iterative approach \cite{log-sum-2,log-sum-3} and similar to \cite{SFW-BS-GC},
the optimization in \eqref{opJ} can be transformed into minimizing the surrogate function as
\begin{align}
\min_{\bm \psi, \bm \tau } \
\underbrace{
- \mathbf y_k^H \mathbf W^H \mathbf P_k (\bm \psi, \bm \tau) \Big(
\mathbf P_k^H (\bm \psi, \bm \tau) \mathbf W \, \mathbf W^H\mathbf P_k (\bm \psi, \bm \tau)
+ \lambda ^{-1} \mathbf D^{(n)}
\Big)^{-1} \mathbf P_k^H (\bm \psi, \bm \tau) \mathbf W\mathbf y_k
+
C(\bm \beta^{(n)})
}_{S_1 (\bm \psi, \bm \tau)}
,
\label{Sl2}
\end{align}
where ${S_1 (\bm \psi, \bm \tau)}$ is as defined,
\begin{align}
\mathbf D^{(n)} & \triangleq
\diag \Bigg\{
\frac{ 1 } { \big| [\bm \beta^{(n)}]_1 \big| ^2 + \epsilon }
,\dots,
\frac{ 1 } { \big| [\bm \beta^{(n)}]_{L_k} \big| ^2 + \epsilon }
\Bigg\}
,
\label{Dn}
\end{align}
and $C(\bm \beta^{(n)})$ is the constant independent of
$\bm \psi$, $\bm \tau$, and $\bm \beta$.
For given $\bm \psi $ and $\bm \tau$, the optimal value of $\bm \beta$ can be obtained as
\begin{align}
\bm \beta^* (\bm \psi, \bm \tau) & = \Big(
\mathbf P_k^H (\bm \psi, \bm \tau) \mathbf W \, \mathbf W^H \mathbf P_k (\bm \psi, \bm \tau)
+ \lambda ^{-1} \mathbf D^{(n)}
\Big)^{-1} \mathbf P_k^H (\bm \psi, \bm \tau) \mathbf W \mathbf y_k
.
\label{beta*}
\end{align}

Although it is difficult to obtain an analytical solution of \eqref{Sl2}, only iterative reduction of $S_1 (\bm \psi, \bm \tau)$ is required in our algorithm. Considering that $S_1 (\bm \psi, \bm \tau)$ is differentiable with respect to both $\bm \psi$ and $ \bm \tau$, gradient descent can be applied in each iteration.
Consequently, \eqref{Sl2} is guaranteed to be non-increasing and a stationary point of $(\bm \psi, \bm \tau)$ will be finally reached.

\subsection{Parameters Selection}
\label{sec:parameter}

In Section~\ref{sec:alg1}, we have discussed initial parameter extraction by the gridless compressive sensing method.
Several parameters need to be carefully selected.

\subsubsection{$\epsilon$}
\label{sec:epsilon}

Consider another form of \eqref{J0},
$J_0 (\bm \beta) \propto \sum_{l}
\log \big( 1 + \big| [\bm \beta]_l \big| ^2 / \epsilon \big)
$,
which tends to the original $\ell_0$-norm in \eqref{op0} as
$\epsilon \to 0$ and indicates that
$\epsilon$ might be set arbitrarily small to make the log-sum most closely resemble the $\ell_0$-norm.
Unfortunately, it is more likely that iterative algorithms will converge extremely slow and get stuck in unenviable values when $\epsilon$ is too small \cite{log-sum-2}.
Moreover, it would induce the tricky issue that optimization objectives and matrices, such as \eqref{Dn}, become ill-conditioned due to the divide-by-zero problem when $[\bm \beta ]_l \simeq 0$.
Therefore, $\epsilon$ should be cautiously chosen to maintain the stability of iterative algorithms.
In the proposed algorithm, instead of remaining fixed,
$\epsilon $ is set a relatively large value at the first iteration and then gradually decreases during the iteration process.
Compared with the fixed $\epsilon $, the slow reduction of $\epsilon $ efficiently speeds up the convergence rate \cite{epsilon-local-minima}.
It should be noted that the optimization (20) or (23) will become a new problem each time $\epsilon$ is updated to a smaller value during iterations. Therefore, the gradual reduction of $\epsilon$ will not influence much on the final solution, which depends upon the final $\epsilon$ at the last iteration.
In the proposed algorithm, $\epsilon $ is initialized to 1 and will reduce to $\epsilon /10$ if
$\| \bm \beta^{(n+1)} - \bm \beta^{(n)} \|_2 < \sqrt{\epsilon}$
until it reaches the preset minimum value, e.g., $10^{-8}$.

\subsubsection{Constant $\lambda$ or adaptive-updating $\lambda^{(n)}$}

The regularization parameter, $\lambda$, determines how much we compromise between the sparsity and the data fitting deviation.
A large $\lambda$ gives heavy weight on the fitting deviation and thus produces a better-fitting solution, which, however, increases the possibility of overestimation.
As $\lambda$ is crucial to the recovery performance and the speed of convergence, how to select it is very important.
To achieve a tradeoff between the sparsity and the data-fitting deviation, we set $\lambda$ as the inverse of the noise variance of vector $\bm \beta$'s elements.
Since the true noise statistics may be unknown,
we select the following
\begin{align}
\lambda^{(n)} & = \max \bigg(
\lambda_0 \cdot \frac{1}{ \|
\mathbf y_k - \mathbf W^H \mathbf P_k (\bm \psi^{(n)}, \bm \tau^{(n)})  \bm \beta^{(n)}
\|_2^2}
\ , \
\lambda_{min} \bigg)
,
\label{lambdan}
\end{align}
where $\lambda_0 $ and $\lambda_{min} $ are two preset constants;
$\bm\psi^{(n)}$ and $\bm\psi^{(n)}$ are the estimated AoAs and path delays at the $n$th iteration.
After each iteration, $\lambda^{(n)}$ dynamically adjusts its value until reaching the threshold $\lambda_{min} $.
The constant, $\lambda_0 $, remains fixed to balance the first item of \eqref{opJ}
and trade off the sparsity and the data-fitting deviation.

\begin{table}
\small
\normalsize
\centering
\caption{Algorithm: Iterative Parameters Extraction for Uplink Channels}
\label{Tab01}

\vspace{-2em} \hrulefill \vspace{-1em}

\begin{flushleft} \justifying \hangafter 1 \hangindent 4.1em
\textbf{Step 1:} \;
Set $n=0$ and $L_k = L_M$.
Initialize $\epsilon$, $\bm \beta^{(0)}$, $\bm \psi^{(0)}$, and $\bm \tau^{(0)}$. Calculate $\lambda^{(n)}$ from \eqref{lambdan}.
\end{flushleft} \vspace{-1.8em}

\begin{flushleft} \justifying \hangafter 1 \hangindent 4.1em
\textbf{Step 2:} \;
For iteration $n$, construct the surrogate function by \eqref{Sl2}.
\end{flushleft} \vspace{-1.8em}

\begin{flushleft} \justifying \hangafter 1 \hangindent 4.1em
\textbf{Step 3:} \;
Optimize the surrogate function to find a new estimate for $\bm \psi^{(n+1)}$ and $\bm \tau^{(n+1)}$ by \\ gradient descend.
\end{flushleft} \vspace{-1.8em}

\begin{flushleft} \justifying \hangafter 1 \hangindent 4.1em
\textbf{Step 4:} \;
Calculate $\bm \beta^{(n+1)}$ by \eqref{beta*} and
update $\lambda^{(n+1)}$ by \eqref{lambdan}.
\end{flushleft} \vspace{-1.8em}

\begin{flushleft} \justifying \hangafter 1 \hangindent 4.1em
\textbf{Step 5:} \;
Calculate $\gamma = \| \bm \beta^{(n+1)} - \bm \beta^{(n)} \|_2$.
If $\gamma < \sqrt{\epsilon}$, then
$\epsilon = \max \{ \epsilon / 10 , \epsilon_{min} \}$.
\end{flushleft}  \vspace{-1.8em}

\begin{flushleft} \justifying \hangafter 1 \hangindent 4.1em
\textbf{Step 6:} \;
For $l$ satisfying $[\bm \beta^{(n+1)}]_l < \beta_{min}$, remove
$[\bm \beta^{(n+1)}]_l$, $[\bm \psi^{(n+1)}]_l$, and $[\bm \tau^{(n+1)}]_l$
from \\ vectors $\bm \beta^{(n+1)}$, $\bm \psi^{(n+1)}$, and $\bm \tau^{(n+1)}$, respectively, and update $L_k$.
\end{flushleft} \vspace{-1.8em}

\begin{flushleft} \justifying \hangafter 1 \hangindent 4.1em
\textbf{Step 7:} \;
Set $n = n+1$.
\end{flushleft}  \vspace{-1.8em}

\begin{flushleft} \justifying \hangafter 1 \hangindent 4.1em
\textbf{Step 8:} \;
Go to Step 2 if $ \gamma < \gamma_T$; otherwise stop and output the results.
\end{flushleft}

\vspace{-1.5em} \hrulefill \vspace{-2em}
\end{table}

\subsubsection{$L_M$}

During the iteration process, the number of channel paths or non-zero channel gains, $L_k$, will gradually decrease from the initial value, $L_M$.
At the $n$th iteration, $[\bm \beta^{(n)}]_l < \beta_{min}$, $\forall l$,  will be deleted from vector $\bm \beta^{(n)}$, where constant $\beta_{min}$ is a preset threshold independent of iteration times.
The corresponding $[\bm \psi^{(n)}]_l$ and $[\bm \tau^{(n)}]_l$ will be meanwhile removed from vector $\bm \psi^{(n)}$ and $\bm \tau^{(n)}$, respectively.
$L_M$ should be set relatively large to make the final results less likely to be trapped in undesirable local minima.
However, if $L_M$ is set excessively large,
it will result in significant computational complexity and slow down convergence.
As a compromise, $L_M$ ought to be large but keep the same order of magnitude of the number of actual channel paths, in terms of different propagation scenarios.

\subsubsection{$\gamma_T$}

We follow the conventional practice to set a hard threshold, $\gamma_T$, as the terminating condition \cite{Eldar-CSbook}. Specifically, the iteration process stops when
$ \| \bm \beta^{(n+1)} - \bm \beta^{(n)} \| < \gamma_T$.

Table~I summarizes the above discussion and presents the concrete steps of the proposed algorithm.

\section{Uplink Channel Estimation}

After obtaining the initial physical parameters,
both uplink and downlink channels can be estimated via a significantly small amount of training.
It relies on the following three facts:
\begin{itemize}
  \item A mobile's physical location changes much slower than the channel variation, which indicates that the coherence times of angles and delays are much longer than that of the channel gains
      \cite{DFT-Xie,Fandian-JSAC}.
      It is reasonable that AoAs and path delays for a user obtained in \emph{initial parameter extraction phase} remain unchanged for a relatively long time,
      depending on its moving speed.
      For example, assume a user is 500 meters away from the BS with the speed of 80~km/h and operate with the bandwidth of $W = 600$~MHz. Within 1~ms, the maximum changes of AoAs and delays will merely be $2.5 \times 10^{-3}$~(deg) and $7.4 \times 10^{-2}$~(ns), respectively, during which hundreds of OFDM blocks will be transmitted. Therefore, only the channel gains are required to be re-estimated or updated for the new coming channel coherent time.

  \item If two users possess different AoAs and path delays, then they can be trained at the same time-frequency band as the BS can distinguish them by the asymptotical \emph{angle-delay orthogonality}, which further reduces the training overhead.
  \item The frequency-insensitive parameters, i.e., AoAs and path delays, can be directly applied in downlink channel estimation.  Therefore, only the channel gains need to be fed back to the BS to reconstruct the downlink channel and the user feedback can then be significantly reduced.
\end{itemize}

\subsection{Uplink User Grouping}
\label{sec:update}

We first propose a criterion for user grouping and scheduling for uplink channel estimation.
In terms of the asymptotical angle-delay orthogonality \cite{SFW-TSP,SFW-CM},
two non-identical paths\footnote{Two paths are defined identical in this paper if they have the identical AoA  and identical delay.} are asymptotically orthogonal.
For finite values of $M$ and $P$ in practice,
we can use the following distance to indicate the orthogonality level between two uplink channels as
\begin{align}
d_U(\mathbf h_{k_1}, \mathbf h_{k_2}) & \triangleq
\min_{\substack
{l_1, l_2}}
\big\| [M \psi_{k_1,l_1},P\eta{\tau}_{k_1,l_1}]^T
 - [M\psi_{k_2,l_2},P\eta{\tau}_{k_2,l_2}]^T
\big\|_2 \ .
\label{dist}
\end{align}
We then assign user $k_1$ and user $k_2$ into the same uplink training group if
$d_U(\mathbf h_{k_1}, \mathbf h_{k_2}) \geq \Omega_U$, where $\Omega_U$ can be deemed as the guard interval.

Denote $G^U$ as the number of uplink groups and $\mathcal G_g^U$ as the set of user indices belonging to group $g \in \{1,\dots,G^U\}$.
Then, multiple users in the same group $g$ are assigned the identical $P$ subcarriers in an OFDM block and transmit the same pilot symbols in the identical time-frequency band, i.e.,
\begin{align}
\left\{
\begin{aligned}
    \mathcal P_k & = \mathcal P_r ,  \quad \forall k,r \in \mathcal G_g^U , g \in \{ 1,\dots,G^U\}
    \\
    \mathcal P_k & \cap \mathcal P_r = \varnothing, \quad \forall k \in \mathcal G_{g_1}^U , r \in \mathcal G_{g_2}^U, g_1 \neq g_2
\end{aligned}
\right.
,
\label{group_rule}
\end{align}
where $\mathcal P_k$ and $\mathcal P_r$ follow the definition in Section~\ref{sec:system}.
By doing this, we can save a large number of pilot resources compared to the initial parameter extraction stage.

Denote the set of pilot subcarriers for group $g$ as
$\mathcal P_g = \{ p_{g,1} ,\dots, p_{g,P} \}$.
The received signal on $\mathcal P_g$ can be written as
\begin{align}
\mathbf y_{g,q,b} = \mathbf W_{q,b}^H \sum_{k \in \mathcal G_g^U}
\mathbf h_{k,q} x_{g,q,b} + \mathbf W_{q,b}^H \mathbf n_{g,q,b}
, \quad
q \in \mathcal P_g
,
\end{align}
where $x_{g,q,b}$ is the pilot symbol at the $q$th subcarrier in the $b$th block and $\mathbf n_{g,q,b} \in \mathbb C^{M\times 1}$ is the corresponding additive Gaussian noise.
Similar to Section~\ref{sec:system}, we have
\begin{align}
\mathbf y_g = \mathbf W_g^H
\sum_{k \in \mathcal G_g^U} \mathbf h_k
+ \mathbf n_g
= \mathbf W_g^H \mathbf P_k \bm \alpha_k +
\sum_{r \in \mathcal G_g^U \backslash \{k\} } \mathbf W_g^H \mathbf P_r \bm \alpha_r
+ \mathbf n_g
,
\label{yg}
\end{align}
where
\begin{align}
{\mathbf W}_g & \triangleq
\left[
\begin{array}{cccc}
\mathbf W_{p_{g,1}} & \mathbf 0 & \cdots & \mathbf 0 \\
\mathbf 0 & \mathbf W_{p_{g,2}}  & \cdots & \mathbf 0 \\
\vdots & \vdots & \ddots & \vdots \\
\mathbf 0 & \mathbf 0 & \cdots & \mathbf W_{p_{g,P}}
\end{array}
\right]
\in \mathbb C^{MP \times N_{\textup{RF}} P T_{\textup{up}}}
\notag
\end{align}
and $
{\mathbf n_g} \triangleq
\Big[
\big( \tilde{\mathbf W}_{p_{g,1}}^H \tilde{\mathbf n}_{p_{g,1}} \big)^T ,\dots,
\big( \tilde{\mathbf W}_{p_{g,P}}^H \tilde{\mathbf n}_{p_{g,P}} \big)^T
\Big]^T
\in \mathbb C^{ N_{\textup{RF}} P T_{\textup{up}} \times 1}
$.

We next design  least-square (LS) and MMSE estimators to update the uplink complex gains and reconstruct the uplink channel.

\subsection{LS Estimator}

In terms of the asymptotical orthogonality among channels of different users in the same training group,
the LS estimate can be immediately obtained as
\begin{align}
\hat{\bm \alpha }_{k,\tu{LS}} & = (\mathbf W_g^H \mathbf P_k)^{\dag} \mathbf y_g
= (\mathbf P_k^H \mathbf W_g \mathbf W_g^H \mathbf P_k)^{-1} \mathbf P_k^H \mathbf W_g \mathbf y_g
\notag \\ & = \bm \alpha _k
+  \sum_{r \in \mathcal G_g^U \backslash \{k\} }
(\mathbf P_k^H \mathbf W_g \mathbf W_g^H \mathbf P_k)^{-1} \mathbf P_k^H \mathbf W_g
\mathbf W_g^H \mathbf P_r \bm \alpha _r
+ (\mathbf P_k^H \mathbf W_g \mathbf W_g^H \mathbf P_k)^{-1} \mathbf P_k^H \mathbf W_g \mathbf n_g
.
\notag
\end{align}
Denote
\begin{align*}
\widetilde{\mathbf p}(\psi_{k,l}, \tau_{k,l}) & \triangleq \Big[
\mathbf a^T (\Xi _{k,l} (0)),
\mathbf a^T (\Xi _{k,l} (\eta)) e^{-j2\pi \eta \tau_{k,l}} \\
 , \dots,
\mathbf a^T & (\Xi _{k,l} ((N_c-1)\eta)) e^{-j2\pi (N_c-1) \eta \tau_{k,l}}
\Big]^T
\in \mathbb C^{MN_c \times 1}
\end{align*}
as the channel basis  for all $N_c$ subcarriers
and define $\widetilde{\mathbf P}_k \triangleq \big[
\widetilde{\mathbf p} (\psi _{k,1} ,\tau_{k,1}),\dots,
\widetilde{\mathbf p} (\psi _{k,L_k} , \tau_{k,L_k})
\big]$.
The uplink channel of user $k$ on all subcarriers can thereupon be
reconstructed as $\widetilde{\mathbf h}_{k,\tu{LS}}  = \widetilde{\mathbf P}_k  \hat{\bm \alpha }_{k,\tu{LS}} \approx \widetilde{\mathbf P}_k \bm \alpha_k$.

\subsection{MMSE Estimator with Reconstructed Covariance Matrices}

Usually, the acquisition of channel statistical information requires long-term average. Nevertheless, equations \eqref{Lambdak} and \eqref{Rku} enlighten us to construct the channel covariance matrix in terms of the physical channel parameters, i.e., AoAs, path delays, and $\bm\Lambda_k$, in a highly efficient way.

Note that $\bm \Lambda_k$ can be calculated from previous averages with much fewer samples than those required in conventional covariance matrix construction. Actually, $\bm \Lambda_k$ can even be replaced by a single estimate of complex gains obtained during  the initial parameter extraction phase.
Such covariance matrices
perform favourably in channel estimation compared with the true ones, as will be seen in Section \ref{sec:simu}.

With the computed covariance matrix, $\mathbf R_k^U$, by equation \eqref{Rku},
we can obtain MMSE estimate of the uplink channel gains from \eqref{yg} as
\begin{align}
\hat{\bm \alpha }_{k,\tu{MMSE}} & =
\bm \Lambda_k \mathbf P_k^H \mathbf W_g \bigg(
\mathbf W_g^H \mathbf P_k \bm \Lambda_k \mathbf P_k^H \mathbf W_g
+  \sum_{r \in \mathcal G_g \backslash \{k\} }
\mathbf W_g^H \mathbf P_r \bm \Lambda_r \mathbf P_r^H \mathbf W_g
 + \sigma _n^2  \mathbf C_{\mathbf n_g}
\bigg)^{-1} \mathbf y_g
\notag\\ & =
\bm \Lambda_k \mathbf P_k^H \mathbf W_g \bigg(
\mathbf W_g^H \sum_{r \in \mathcal G_g }
\mathbf R_r \mathbf W_g + \sigma _n^2  \mathbf C_{\mathbf n_g}
\bigg)^{-1} \mathbf y_g
,
\label{alpha_MMSE}
\end{align}
where
\begin{align}
\mathbf C_{\mathbf n_g} & =
\left[
\begin{array}{cccc}
\tilde{\mathbf W}_{p_{g,1}}^H \tilde{\mathbf W}_{p_{g,1}} & \mathbf 0 & \cdots & \mathbf 0 \\
\mathbf 0 & \tilde{\mathbf W}_{p_{g,2}}^H \tilde{\mathbf W}_{p_{g,2}}  & \cdots & \mathbf 0 \\
\vdots & \vdots & \ddots & \vdots \\
\mathbf 0 & \mathbf 0 & \cdots & \tilde{\mathbf W}_{p_{g,P}}^H \tilde{\mathbf W}_{p_{g,P}}
\end{array}
\right]
\in \mathbb C^{N_{\textup{RF}} P T_{\textup{up}} \times N_{\textup{RF}} P T_{\textup{up}}}
\notag
\end{align}
and the uplink channel of user $k$ on all $N_c$ subcarriers can thereupon be updated as
\begin{align}
\widetilde{\mathbf h}_{k,\tu{MMSE}} & = \widetilde{\mathbf P}_k  \hat{\bm \alpha }_{k,\tu{MMSE}}
 = \widetilde{\mathbf P}_k
\bm \Lambda_k \mathbf P_k^H \mathbf W_g \bigg(
\mathbf W_g^H \sum_{r \in \mathcal G_g }
\mathbf R_r \mathbf W_g + \sigma _n^2  \mathbf C_{\mathbf n_g}
\bigg)^{-1} \mathbf y_g
.
\label{hk_LMMSE}
\end{align}

\section{Precoding Design at BS and Downlink Channel Estimation}
\label{sec:dl}

For time-division duplex (TDD) systems, downlink channels can be immediately obtained via the reciprocity between uplink and downlink, which is not the case for FDD systems. We here, design a novel downlink channel estimation strategy for FDD systems with significantly low training overhead and limited user feedback by exploiting the \emph{AoA-delay reciprocity} \cite{DFT-Xie,SFW-TSP} and the sparsity of mmWave massive MIMO channels, where the \emph{beam squint effect} is carefully considered.

In conventional cases without considering beam squint, for each multipath component, one can simply use a single RF chain to generate a beam pointing towards the specified direction.
Considering the beam squint effect, a beam should be generated by frequency-dependent beam-steering vectors over different subcarriers,
which cannot be achieved by a single RF chain since the analog precoders, i.e., the phase setup in phase shifters, are generally fixed and constant during one OFDM block.
Ignoring beam squint will prevent the signals in certain frequencies from reaching the specified users.
To address this issue, we propose to utilize several RF chains to cooperatively generate the frequency-dependent beamforming vector across different subcarriers.

In this section, after introducing the downlink channel model and user grouping, we present our analog and digital precoder design at the BS to address the beam squint issue and then discuss downlink channel estimation.

\subsection{Downlink Channel Model and User Grouping}

Denote the downlink carrier frequency and wavelength as $f_c^D$ and $\lambda_c^D = c / f_c^D$, respectively.
The downlink channel can be immediately formulated as
\begin{align}
\mathbf h_k^D & = \mathbf P_k(\bm \psi_k^D, \bm \tau_k) \bm \alpha_k^D
,
\label{hD}
\end{align}
where $\bm \psi_k^D = [\psi_{k,1}^D,\dots,\psi_{k,L_k}^D]^T$ can be directly computed from the uplink version extracted in initial parameter extraction phase as
\begin{align}
\psi_{k,l}^D & = \frac{d \sin \vartheta _{k,l}}{\lambda _c^D}
= \frac{f_c^D}{f_c} \frac{d \sin \vartheta _{k,l}}{\lambda _c}  = \frac{f_c^D}{f_c} \psi_{k,l}
.
\label{psiD}
\end{align}

Since users are unaware of path delays and do not necessarily synchronize with each other,
we propose to group users only in terms of AoAs. Accordingly, the distance between two downlink channels  is defined as
\begin{align}
d_D(\mathbf h_{k_1}^D, \mathbf h_{k_2}^D) & \triangleq
\min_{\substack
{l_1, l_2}}
\big| M \psi_{k_1,l_1}^D,
 - M\psi_{k_2,l_2}^D
\big|^2.
\label{dist_D}
\end{align}
Similarly, we preset the corresponding guard distance, $\Omega_D$, such that user $k_1$ and user $k_2$ are assigned into the same downlink group if $d_D(\mathbf h_{k_1}^D, \mathbf h_{k_2}^D) \geq \Omega_D$.
Due to the limited RF chains in hybrid digital/analog architecture transceivers, users from all spatial directions might not able to be completely covered within a single OFDM block. We employ $T_{\textup{dl}}$ successive OFDM blocks for the downlink channel estimation.
As a result, multiple users in the same group will be trained at the identical $P$ subcarriers within $T_{\textup{dl}}$ OFDM block.

Denote $G^D$ as the number of downlink groups and $\mathcal G_g^D$ as the set of user indices belonging to group $g \in \{1,\dots,G^D\}$.
Denote $\mathbf h_{k,q}^D$ as the downlink channel at $q$th subcarrier of user $k$.
In the $g$th group, the received signal of the $k$th user at the $q$th subcarrier from $M$ antennas at the BS in the $b$th block can be expressed as
\begin{align}
y_{k,q,b} = (\mathbf h_{k,q}^D)^H \mathbf F_{\textup{RF},t} \mathbf F_{\textup{BB},q,b}
\mathbf s_{g,q,b}
+ n_{k,q,b}
,
\end{align}
where $\mathbf s_{g,q,b} \in \mathbb C^{N_{\textup{RF}} \times 1}$ is the pilot symbol for the $g$th group, at the $q$th subcarrier, and in the $b$th block, and $n_{k,q,b} $ is the corresponding additive Gaussian noise.
Making a summation of the received signal over $T_{\textup{dl}}$ blocks for each subcarrier yields
\begin{align}
\tilde{y}_{k,q} = \sum_{t = 1}^{T_{\textup{dl}}} y_{k,q,b} =
\mathbf h_{k,q}^H
{\mathbf F}_{\textup{RF}} \mathbf F_{\textup{BB},q}
\mathbf s_{g,q} + \tilde{n}_{k,q}
,
\end{align}
where
$\mathbf F_{\textup{RF}} \triangleq \big[
\mathbf F_{\textup{RF},1} , \mathbf F_{\textup{RF},2} ,\dots, \mathbf F_{\textup{RF},T_{\textup{dl}}}
\big] \in \mathbb C^{M \times N_{\textup{RF}} T_{\textup{dl}}}$,
\begin{align}
\mathbf F_{\textup{BB},q} \triangleq
\left[
\begin{array}{cccc}
\mathbf F_{\textup{BB},q,1} & \mathbf 0 & \cdots & \mathbf 0 \\
\mathbf 0 & \mathbf F_{\textup{BB},q,2}  & \cdots & \mathbf 0 \\
\vdots & \vdots & \ddots & \vdots \\
\mathbf 0 & \mathbf 0 & \cdots & \mathbf F_{\textup{BB},q,T_{\textup{dl}}}
\end{array}
\right]
\in \mathbb C^{N_{\textup{RF}} T_{\textup{dl}} \times N_{\textup{RF}} T_{\textup{dl}}}
,
\label{FBBq}
\end{align}
$\mathbf s_{g,q} \triangleq  \big[\mathbf  s_{g,q,1}^T ,\dots, \mathbf s_{g,q,T_{\textup{dl}}}^T \big]^T \in \mathbb C^{N_{\textup{RF}} T_{\textup{dl}} \times 1}$,
and $\tilde{n}_{k,q}  \triangleq  \sum_{b = 1}^{T_{\textup{dl}}} n_{k,q,b}$.

\subsection{Analog and Digital Precoder Design with Beam Squint}

To design the analog precoder, we first collect all required steering vectors into a set as
\begin{align}
\mathcal A_0 & = \big\{ \mathbf a \big(
\Xi_{k,l}^D \big( (q-1)\eta \big) \big)
\, \big| \,
k\in\{ 1,\dots,K\},
l\in\{ 1,\dots,L_k \},
q\in\{ p_{k,1} ,\dots, p_{k,P} \}
\big\},
\end{align}
where
\begin{align}
\Xi_{k,l}^D (f) \triangleq  \left(1+\frac{f}{f_c^D}\right) \psi _{k,l}^D
.
\end{align}
Then, we employ the following mutually-orthogonal steering vectors,
\begin{align}
\mathcal A_{\textup{orth}} & = \bigg\{
\mathbf a \bigg(
\frac{\lfloor M \cdot \Xi_{k,l}^D \big( (q-1)\eta \big)
\rfloor }{M} \bigg)
\, \bigg| \,
k\in\{ 1,\dots,K\},
l\in\{ 1,\dots,L_k \},
q\in\{ p_{k,1} ,\dots, p_{k,P} \}
\bigg\}
\notag \\ & \bigcup
\bigg\{
\mathbf a \bigg(
\frac{\lceil M \cdot \Xi_{k,l}^D \big( (q-1)\eta \big)
\rceil }{M} \bigg)
\, \bigg| \,
k\in\{ 1,\dots,K\},
l\in\{ 1,\dots,L_k \},
q\in\{ p_{k,1} ,\dots, p_{k,P} \}
\bigg\}
,
\notag\label{Aorth}
\end{align}
where $\lfloor x \rfloor$ denotes the maximum integer that is not bigger than $x$ and
$\lceil x \rceil$ denotes the minimum integer that is not smaller than $x$.
Without loss of generality, assume that $\frac{| \mathcal A_{\textup{orth}} |}{N_{\textup{RF}}}$ is an integer.
Since $| \mathcal A_{\textup{orth}} | \leq M $, the downlink training process is proposed to be completed in $T_{\textup{dl}} = \frac{| \mathcal A_{\textup{orth}} |}{N_{\textup{RF}}}  \leq \frac{M}{N_{\textup{RF}}}$ blocks.
We expect to use the linear combination of the steering vectors in $\mathcal A_{\textup{orth}}$ to approximately generate all steering vectors in $\mathcal A_0$.

To this end,  the analog precoder, $\mathbf F_{\textup{RF}} \in \mathbb C^{M \times |\mathcal A_{\textup{orth}}|} $, is designed by stacking all steering vectors in $\mathcal A_{\textup{orth}}$ into a matrix with each column being one element (one steering vector) in $\mathcal A_{\textup{orth}}$.
The digital precoder at the $q$th subcarrier,
$\mathbf F_{\textup{BB},q} $,
is designed as a diagonal matrix with
\begin{align}
\textup{diag}(\mathbf F_{\textup{BB},q}) = \mathbf F_{\textup{RF}}^{\dag}
\bigg( \sum_{k \in \mathcal G_g^D} \mathbf B_{k,q}
\bigg) \mathbf c_q
=
(\mathbf F_{\textup{RF}}^H \mathbf F_{\textup{RF}})^{-1}
\mathbf F_{\textup{RF}}^H
\bigg( \sum_{k \in \mathcal G_g^D} \mathbf B_{k,q}
\bigg) \mathbf c_q
,
\end{align}
where $\mathbf B_{k,q} = [\bm \rho_{k,q,1} ,\dots, \bm \rho_{k,q,L_k} ] \in \mathbb C^{M \times L_k}$ is the spatial beamforming vector at the $q$th subcarrier for the $k$th user to eliminate the inter-user interference among users in the same group with
\begin{align}
\bm \rho_{k,q,l} & = \frac{e^{-j2\pi (q-1) \eta \tau_{k,l}} }{M}
\mathbf a \big(
\Xi_{k,l}^D \big( (q-1)\eta \big) \big)
\end{align}
and
$\mathbf c_q$ comes from matrix
$
\mathbf C  =  [\mathbf c_1 ,\dots, \mathbf c_{P} ] \in \mathbb C^{L_k \times P}
$
with mutually-orthogonal rows ($P \geq \max_{k\in\{ 1,\dots,K\}} L_k $).
By doing this, different subcarriers are deployed frequency-dependent beamforming vectors to address the beam squint effect.

\subsection{Downlink Channel Estimation with LS or MMSE estimator}

Denote $\mathbf y_k  \triangleq  [ y_{k,1} ,\dots, y_{k,P} ]^H \in \mathbb C^{P \times 1}$ and assume that the pilot symbols in pilot subcarriers are all 1's for simplicity.
Utilizing the asymptotical orthogonality among different user channels in the same group,
the LS estimate of downlink complex gains can be obtained as
\begin{align}
\hat{\bm \alpha }_{k,\tu{LS}} & =
\big(\mathbf C^H \big)^{\dagger} \mathbf y_k.
\label{alphaD_LS}
\end{align}

Analogously, we can construct the downlink channel covariance matrix
$\mathbf R_k^D$
from \eqref{Rku}, wherein ${\mathbf P_k} (\bm \psi_k, \bm \tau_k)$ is replaced by ${\mathbf P_k} (\bm \psi_k^D, \bm \tau_k)$.
$\bm \Lambda_k$ can also be calculated from the average of previous estimated gains or even replaced by a single estimate of complex gains obtained in the initial uplink parameter extraction phase.
Defining $\mathbf P_k^D \triangleq {\mathbf P_k} (\bm \psi_k^D, \bm \tau_k)$ for notational simplicity, the MMSE estimate of the downlink complex gains can be readily determined as
\begin{align}
\hat{\bm \alpha }_{k,\tu{MMSE}} & =
\bm \Lambda _k (\mathbf P_k^D)^H \bm \Sigma_g
(\bm \Sigma_g^H \mathbf P_k^D \bm \Lambda _k (\mathbf P_k^D)^H \bm \Sigma_g
  + \sigma_n^2 T_{\textup{dl}} \mathbf I_P)^{-1} \mathbf y_k
\notag\\ & =
\bm \Lambda _k (\mathbf P_k^D)^H \bm \Sigma_g
(\bm \Sigma_g^H \mathbf R_k^D \bm \Sigma_g
  + \sigma_n^2 T_{\textup{dl}} \mathbf I_P)^{-1} \mathbf y_k
,
\end{align}
where
\begin{align}
\bm \Sigma_g & \triangleq
\left[
\begin{array}{cccc}
\mathbf F_{\textup{RF}} \, \textup{diag}\{ \mathbf F_{\textup{BB},1} \} & \mathbf 0 & \cdots & \mathbf 0 \\
\mathbf 0 & \mathbf F_{\textup{RF}} \, \textup{diag}\{ \mathbf F_{\textup{BB},2} \}   & \cdots & \mathbf 0 \\
\vdots & \vdots & \ddots & \vdots \\
\mathbf 0 & \mathbf 0 & \cdots & \mathbf F_{\textup{RF}} \, \textup{diag}\{ \mathbf F_{\textup{BB},P} \}
\end{array}
\right]
\in \mathbb C^{MP\times P}.
\end{align}

Finally, the downlink channel of user $k$ on all $N_c$ subcarriers can be updated and reconstructed as
$
\widetilde{\mathbf h}_{k}^D  = \widetilde{\mathbf P}_k^D  \hat{\bm \alpha }_{k,\tu{LS/MMSE}}
$
, where
$\widetilde{\mathbf P}_k^D \triangleq \big[
\widetilde{\mathbf p} (\psi _{k,1}^D ,\tau_{k,1}),\dots,
\widetilde{\mathbf p} (\psi _{k,L_k}^D , \tau_{k,L_k})
\big]$.

We summarize the proposed downlink channel estimation strategy as follows\footnote{The update of uplink channel gains and the reconstruction of uplink channels can simultaneously proceed during Step 2 and Step 3 at a different frequency band.}:
\begin{itemize}
  \item \textbf{Step 1:} \emph{Initial uplink parameter extraction.} Users send orthogonal training pilots to the BS. The BS then applies the proposed algorithm in Table I to extract the physical parameters of each channel path.
  \item \textbf{Step 2:} \emph{Downlink channel gains update and feedback}. The BS sends the downlink pilots to users\footnote{If the MMSE estimator is applied, then the BS also need send the extracted AoAs and path delays to users.}, which are grouped and can be efficiently trained in the same time-frequency band.
      Users estimate downlink channel gains via LS or MMSE estimators proposed in Section~\ref{sec:dl} and then feed them back to the BS.
  \item \textbf{Step 3:} \emph{Downlink channel reconstruction.} The BS reconstructs the downlink channels from the obtained AoAs and path delays in Step 1 as well as the updated downlink channel gains in Step 2.
\end{itemize}

\section{Simulation Results}
\label{sec:simu}

In this section, we present numerical results to demonstrate the necessity of carefully treating the beam squint effect and validate the proposed approaches under practical mmWave massive MIMO system configurations.
The BS is equipped with a ULA whose antenna spacing is half of the downlink carrier wavelength.
The uplink and downlink carrier frequencies are $f_c = 26$~GHz and $f_c^D = 28$~GHz, respectively.
Single-antenna users are uniformly distributed throughout the cell. The AoAs are assumed to be uniformly distributed in $(-\pi/2 , \pi/2)$.
The channel of each user consists of one LoS path and $0 \sim 5$ non-LoS components.
The delay of each multipath component is uniformly drawn from $0 \sim 300$~ns.
We employ the absolute mean-squared error (AMSE) and the normalized mean-squared error (NMSE) as performance indicators.

First, we focus on initial parameter extraction and compare our proposed approach with the conventional off-grid compressive sensing method \cite{CE5-wideband-atheta-Fang-tensor-simu}, which considers the frequency selectivity but simply ignores the beam squint effect.
\begin{figure}
\centering
\includegraphics[width=90mm]{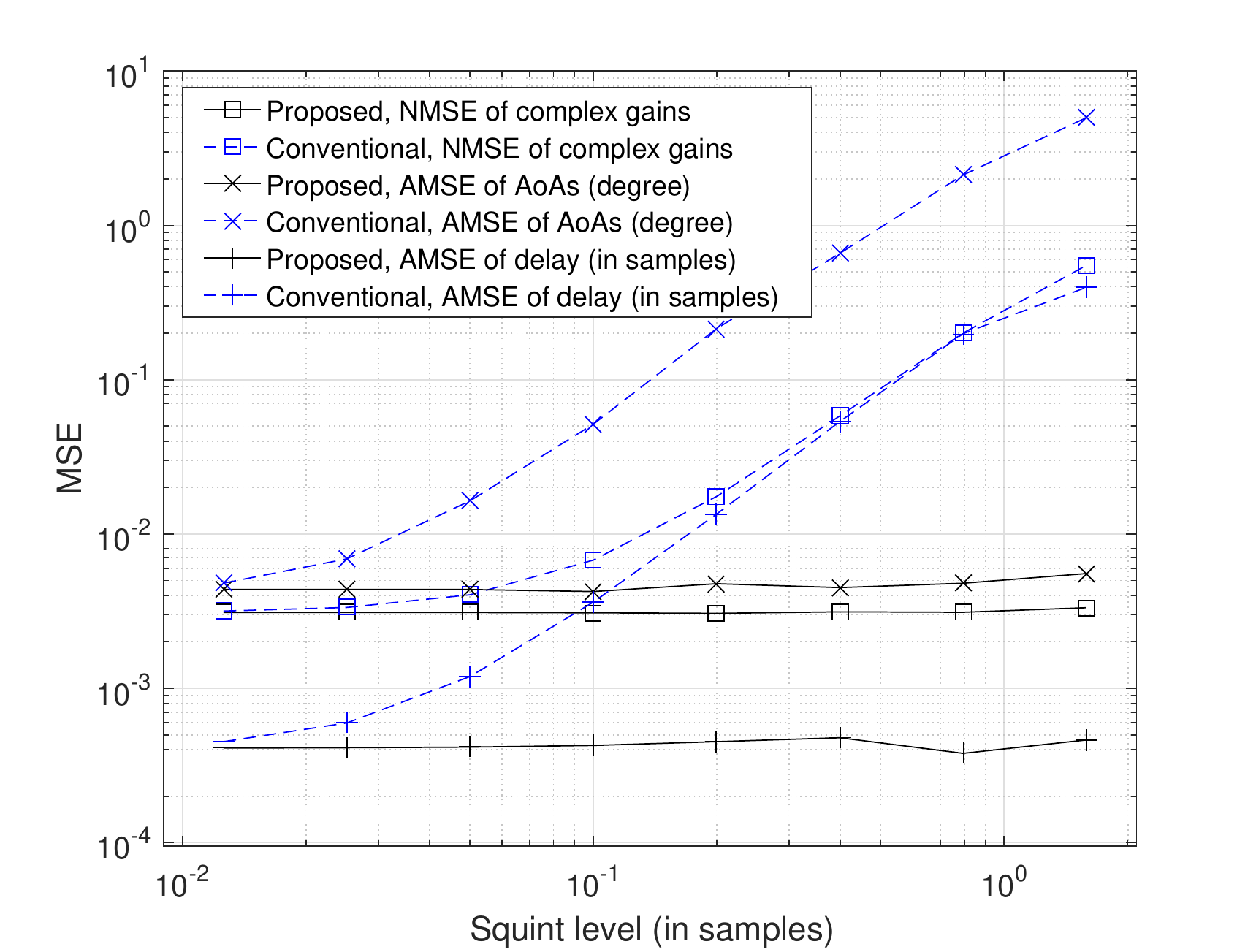}
\caption{MSE of initial parameters extraction versus beam squint level in samples, with $M = 64$, $P=12$, $N_{\textup{RF}}=4$, and $T_{\textup{up}}=12$.}
\label{figBS}
\end{figure}
Note that users do not share the pilot subcarriers and are trained individually at this initial parameter extraction stage.
The squint level in Fig.~\ref{figBS} is depicted by the maximum physical propagation delay in samples,
$\frac{1}{T_s}\tau_{max}^{prop}$,
since it also represents the maximum beam squint along angular indices over subcarriers
according to \eqref{beam_squint_index} in Theorem~1.
The AMSEs of the estimated AoA and path delay are defined as
\begin{align}
\tu{AMSE}_{\vartheta}  = \mathbb E \big\{ | \hat {\vartheta } - \vartheta |^2 \big\}
\quad \textup{and} \quad
\tu{AMSE}_{\tau}  = \frac{1}{T_s} \mathbb E \big\{ | \hat {\tau } - \tau |^2 \big\}
,
\end{align}
respectively, and
the NMSE of a complex gain is defined as
\begin{align}
\tu{NMSE}_{\alpha}  = \frac{\mathbb E \big\{ | \hat {\alpha  } - \alpha  |^2 \big\}
}{
\mathbb E \big\{ |  \alpha  |^2 \big\}
}
.
\end{align}
From Fig.~\ref{figBS}, with the growing squint level, the conventional compressed sensing method fails to extract path AoAs and delays
 due to channel model mismatch while the proposed algorithm maintains its consistent performance.
Actually,
the conventional method implicitly assumes that the observed AoAs at different subcarriers are the same. When beam squint renders the observed AoAs frequency-dependent, the extracted AoAs by such approaches would become a sort of average in certain sense and unpredictable, which even brings the counterproductive effect, i.e., the performance gets worse with more antennas and/or larger bandwidths, as will be shown in the subsequent numerical results.
Note that the squint level of 0.8 corresponds to the bandwidth $W=660$~MHz under current system parameters.

\begin{figure}
\centering
\includegraphics[width=90mm]{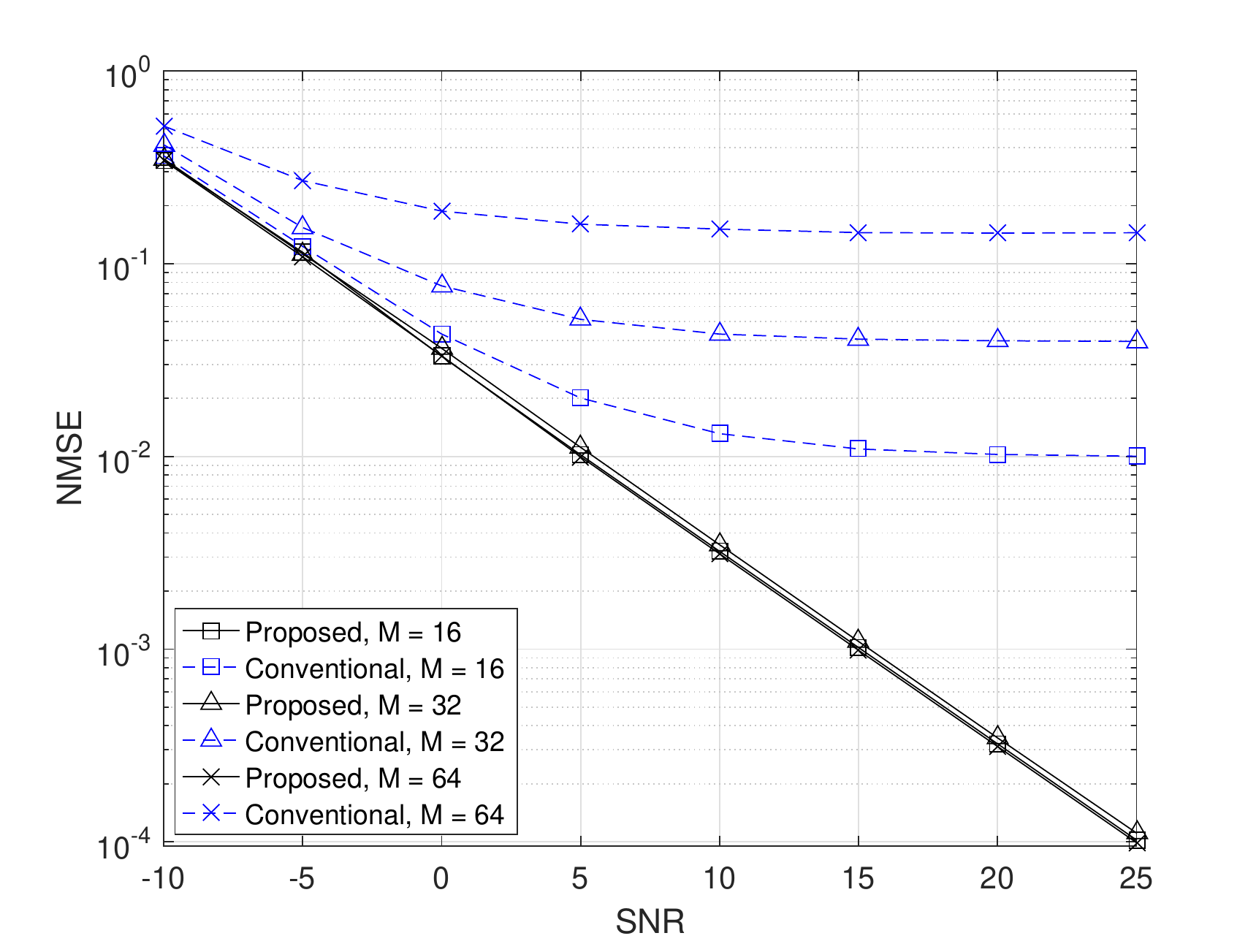}
\caption{NMSE of MMSE estimation for uplink channels versus SNR under different numbers of BS antennas, with $W=600$~MHz, $P=12$, $N_{\textup{RF}}=4$, and $T_{\textup{up}}=12$.}
\label{figM}
\end{figure}

\begin{figure}
\centering
\includegraphics[width=90mm]{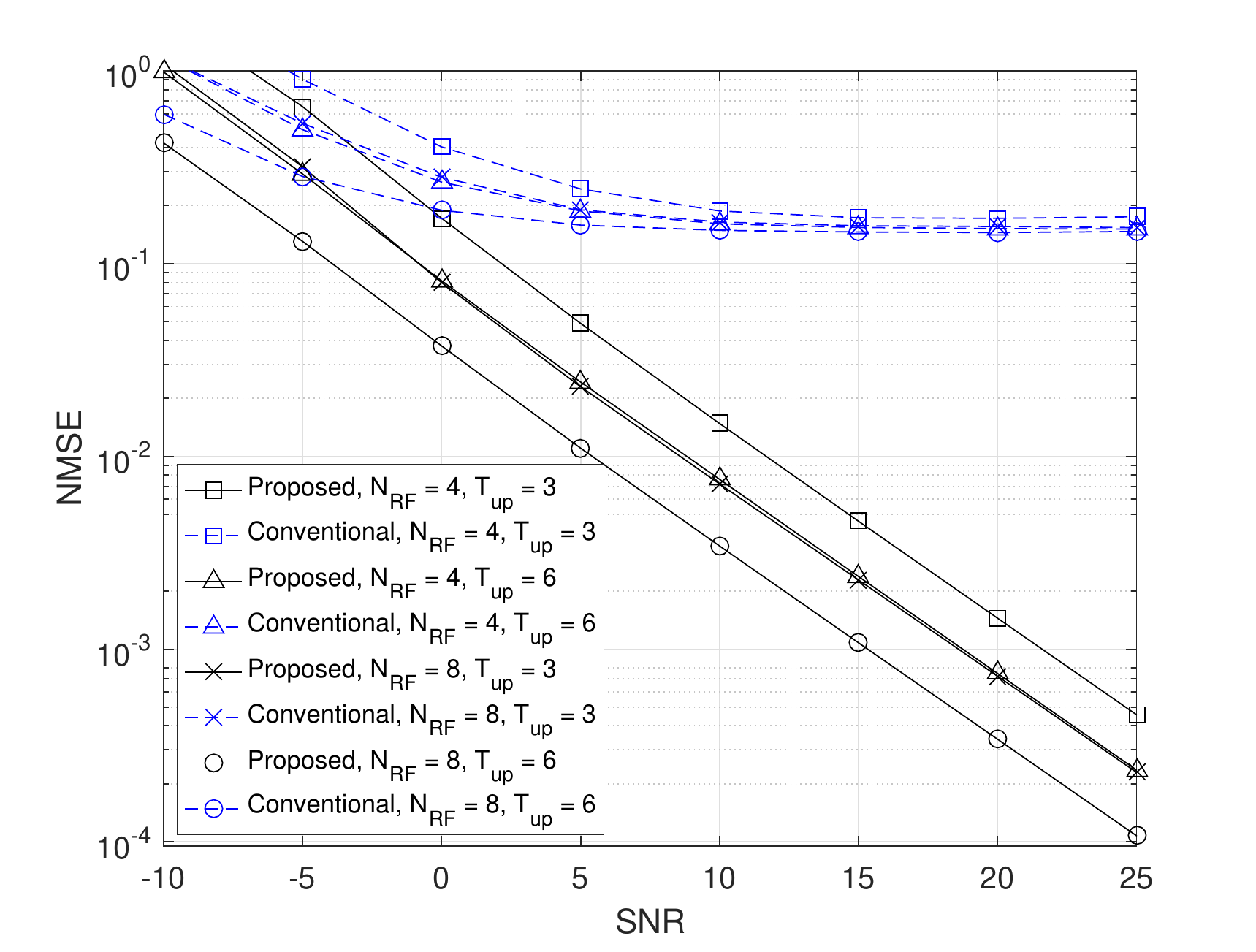}
\caption{NMSE of MMSE estimation for uplink channels versus SNR under different numbers of RF chains and OFDM blocks, with $M=64$, $W=600$~MHz, and $P=12$.}
\label{figNRF}
\end{figure}

\begin{figure}
\centering
\includegraphics[width=90mm]{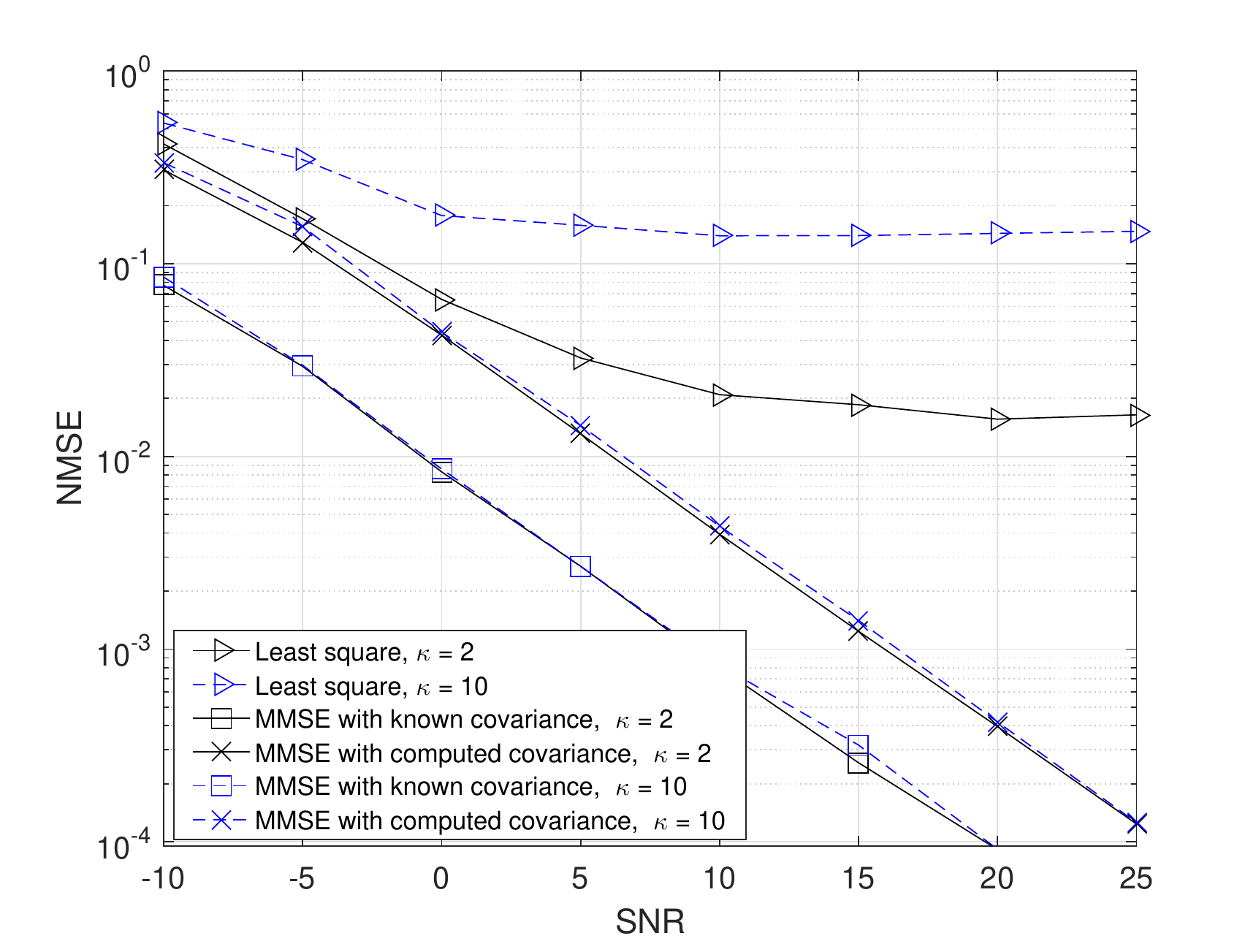}
\caption{NMSE of LS and MMSE estimations for uplink channels versus SNR, with $M = 32$, $P=12$, $W = 600$~MHz, $N_{\textup{RF}}=4$, $T_{\textup{up}}=12$, and $\kappa = 2$~and~$10$ respectively. }
\label{figUL1}
\end{figure}

\begin{figure}
\centering
\includegraphics[width=90mm]{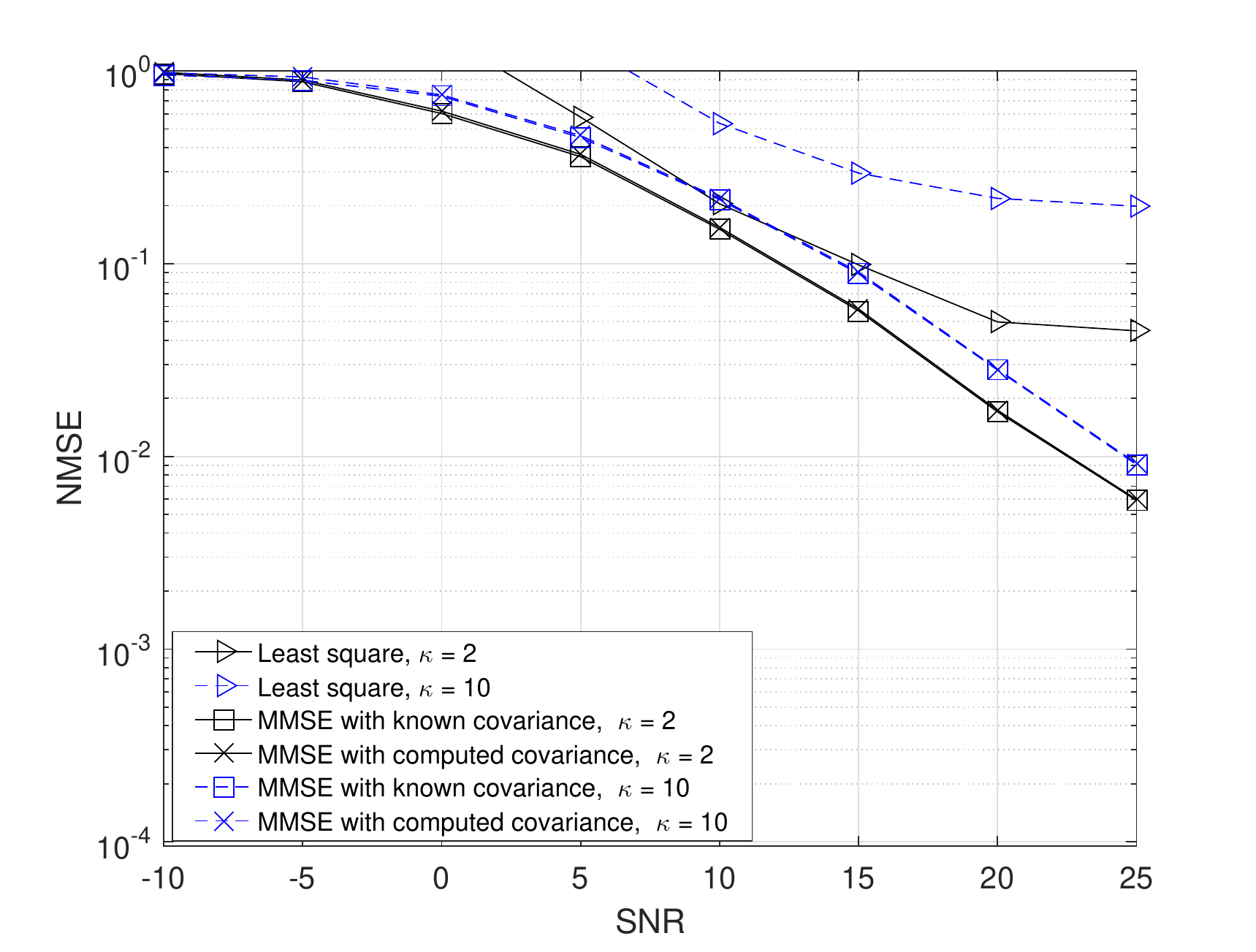}
\caption{NMSE of downlink channel estimation versus SNR, with $M = 32$, $P=12$, $W = 600$~MHz, $N_{\textup{RF}}=4$, and $\kappa = 2$~and~$10$ respectively. }
\label{figDL}
\end{figure}

\begin{figure}
\centering
\includegraphics[width=90mm]{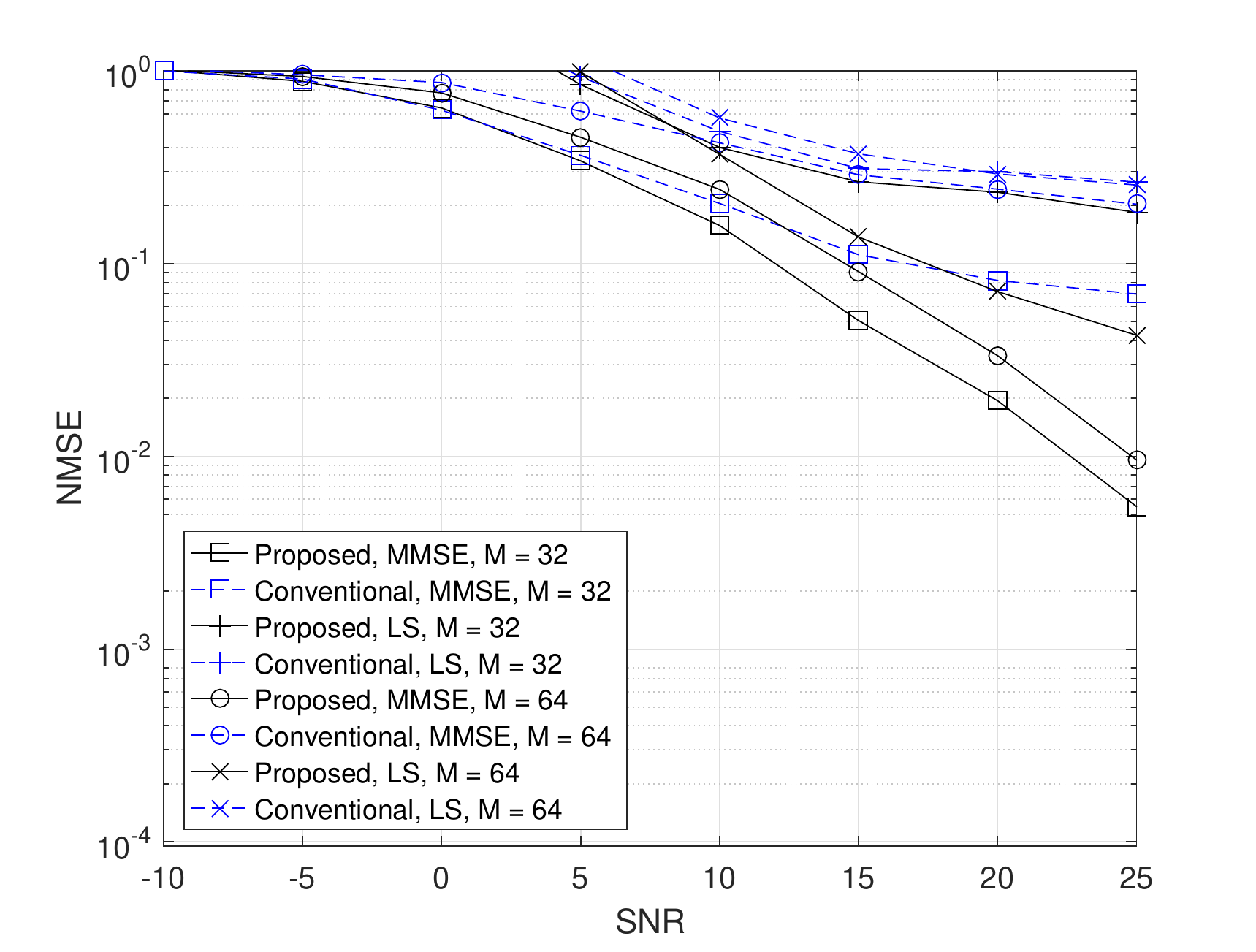}
\caption{NMSE of downlink channel estimation versus SNR, with $M=32$ and $64$, respectively, $P=12$, $W = 600$~MHz, $N_{\textup{RF}}=8$, and $\kappa = 10$. }
\label{figDLM}
\end{figure}

In Fig.\ref{figM}--\ref{figDLM}, we employ the NMSE of estimated channels,
\begin{align}
\tu{NMSE}_{\mathbf h}  = \frac{\mathbb E \big\{ | \hat {\mathbf h} - \mathbf h|_2^2 \big\}
}{
\mathbb E \big\{ |  \mathbf h  |_2^2 \big\}
}
,
\end{align}
as the performance indicator.

Fig.~\ref{figM} depicts the NMSE of MMSE estimation for uplink channels versus the received signal-to-noise ratio (SNR).
In this figure, we consider the single-user scenario to exclusively illustrate the effect of beam squint.
The user is assigned $P = 12$ pilot subcarriers.
The numbers of the BS antennas are set $M = 16$, $32$, and $64$, respectively.
The bandwidth is set $W = 600$~MHz.
The number of RF chains is $N_{\textup{RF}}=4$ and the employed OFDM blocks is $T_{\textup{up}}=12$.
With more BS antennas, the proposed approach maintains remarkable estimation performance. However, the approach ignoring the beam squint effect \cite{CE5-wideband-atheta-Fang-tensor-simu} suffers from severe performance degradation and error floors, increasing with the number of the BS antennas. Actually, more antennas bring counterproductive effect for the conventional approaches since the beam squint effect turns severer.

Compared with Fig.~\ref{figM}, Fig.~\ref{figNRF} fixes the number of antennas and alters the number of RF chains and the employed OFDM blocks.
With more RF chains and OFDM blocks, the proposed approach provides the better estimation performance while the conventional approach cannot benefit from it due to the significantly high error floors.

In the subsequent figures, we investigate the \emph{multi-user scenario}. Fig.~\ref{figUL1} compares the proposed LS and MMSE estimators in uplink channel estimation.
MMSE estimations are achieved by the true covariance and the constructed covariance from initial parameter extraction stage, respectively.
The number of the BS antennas and the shared pilot subcarriers are set
$M = 32$ and $P = 12$, respectively.
The uplink guard interval for eliminating inter-user interference in our simulation is $\Omega_U =5$.
To investigate the impact of frequency reuse or sharing, we limit the maximum number of users in one group, $\kappa \triangleq \max\{|\mathcal G_g^U|\}$, as 2 and 10, respectively.
From Fig.~\ref{figUL1}, the NMSEs of the MMSE estimation with the known covariance and the computed one perform similarly with a performance gap, which results from the estimation deviations of AoAs and path delays.
Therefore, \emph{for channel estimation}, the computed covariance matrices constructed by physical channel parameters provide an excellent approximation of the true ones.
Although the expectation of path power, $\bm \Lambda_k$ defined in \eqref{Lambdak}, is estimated by only one realization and not that accurate, such inaccuracy does not impact much on the final channel estimation result.
Instead, what really matters in a covariance matrix is channel subspace that depends exclusively upon AoAs and path delays.
Since MMSE estimation can effectively eliminate the inter-user interference caused by the frequency reuse, it performs better than LS estimation, as shown in Fig.~\ref{figUL1}, especially when the number of users is large.

The proposed downlink channel estimation presents the similar pattern, as shown in Fig.~\ref{figDL}.
The guard interval for the downlink case is $\Omega_D = 0.4$.
Compared to the uplink case, the MMSE performance gap between the known and the computed covariances is much smaller and ignorable, which further validates the effectiveness of computed covariance matrices.
Since only the angular information is exploited for downlink user grouping and scheduling, the LS estimator, which does not consider the inter-user interference, performs poorly when there are many frequency-reuse users while the MMSE estimator consistently maintains the remarkable performance.

Fig.~\ref{figDLM} shows the downlink channel estimation NMSEs of the proposed channel model and the conventional one that ignores the beam squint effect. The received power at user side is normalized such that users receive the same power under different numbers of antennas, $M$.
For the LS estimator, the proposed approach achieves the better performance with the increase of $M$ since more antennas can provide more accurate beamforming to effectively alleviate the inter-user interference. However, the conventional one performs even worse with more antennas as
it fails to extract the precise AoAs due to beam squint.
For the same reason, MMSE estimators based on the conventional channel model behave even worse than LS due to providing the inaccurate channel subspace information.

\begin{figure}
\centering
\includegraphics[width=90mm]{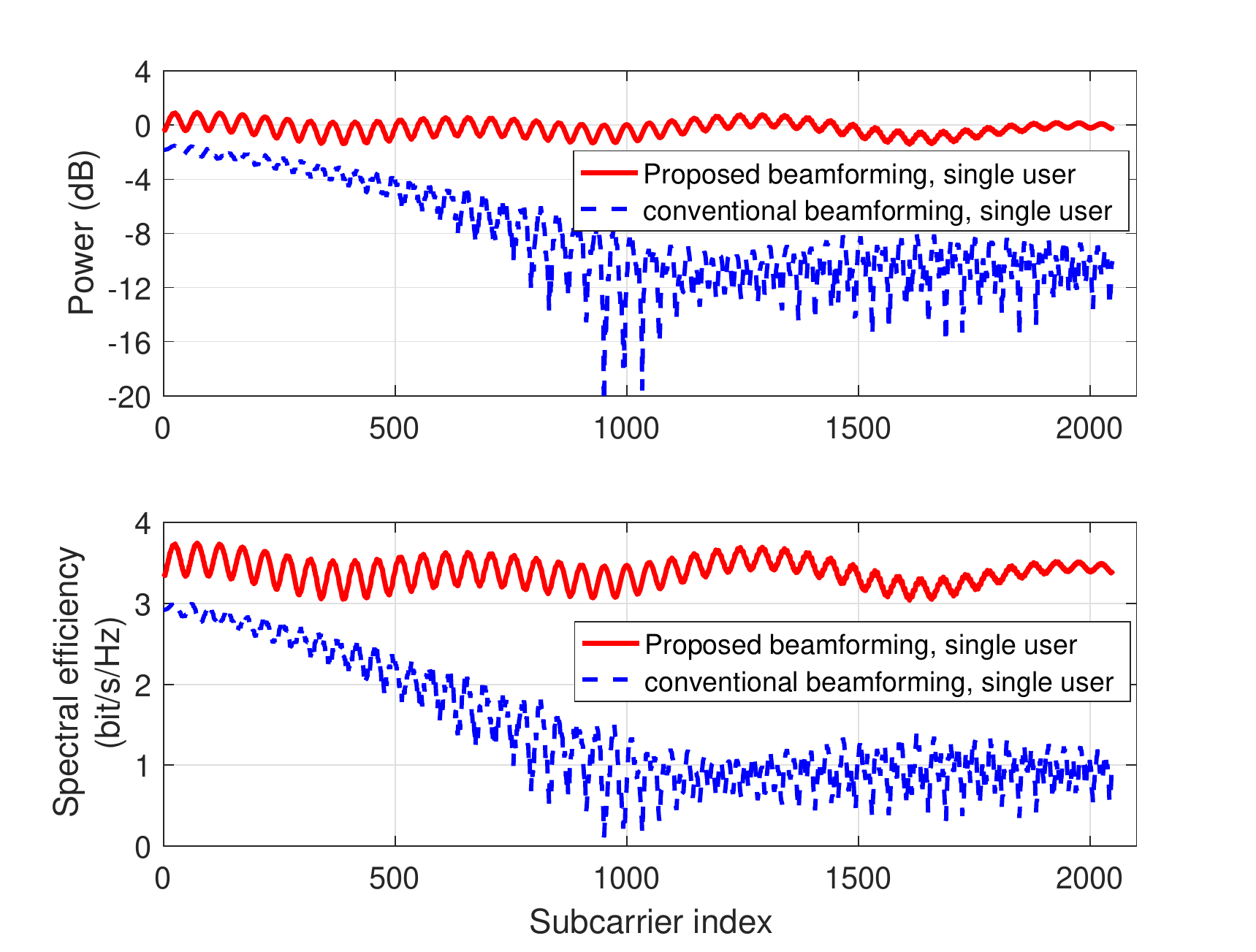}
\caption{Received power and spectral efficiency of a single user in different subcarriers under the proposed and the conventional beamforming strategies in a three-path mmWave channel.}
\label{figBER}
\end{figure}
We here give an example of how beam squint affects the achievable rate at different frequencies as to a specific user. In this example, the channel consists of one LoS path with AoA~$75$\degree  and two NLoS paths with AoAs~$25$\degree and $-20$\degree, respectively. The BS is equipped with a 128-antenna ULA, operating at 28~GHz with bandwidth of 900~MHz.
The SNR at the user side is 10~dB.
As the conventional approach does not consider the beam squint effect, different subcarriers point towards different physical directions and many subcarriers squint to derailed directions. Consequently, the energy from the squinted subcarrier cannot arrive at specified users, which are likely to fall into sidelobes or even null points of the BS's radiation pattern in certain subcarriers.
In Fig.~\ref{figBER}, the energy difference between subcarriers under the conventional beamforming accounts for up to 18~dB, which causes the momentous discrepancy of the spectral efficiencies over frequencies.
To address this issue, the proposed beamforming scheme employs several cooperative RF chains and utilizes the digital precoders to make users locate at the mainlobes in all subcarriers as far as possible.

\section{Conclusions}
In this paper, we investigate the beam squint effect and propose a wideband channel estimation strategy for FDD mmWave massive MIMO systems with hybrid precoding.
We have first presented a channel model with physical parameters and frequency-dependent steering vectors to depict the beam squint effect.
A super-resolution compressive sensing-based approach has been developed to extract the frequency-sensitive parameters (AoAs and path delays) and the frequency-insensitive ones (complex gains).
Then, we have proposed the uplink and downlink channel estimation strategies, which can estimate and update channels
with a significantly small amount of training and user feedback in FDD systems.
Finally, numerical results have demonstrated the superiority of the proposed channel model and channel estimation strategies over the algorithms based on the conventional MIMO models under general mmWave system configurations.

\linespread{1.28}

\end{document}